# Noisy intermediate-scale quantum computers

Bin Cheng,[1,2,3] Xiu-Hao Deng,[1,2,3] Xiu Gu,[1,2,3] Yu He,[1,2,3] Guangchong Hu,[1,2,3] Peihao Huang,[1,2,3] Jun Li,[1,2,3] Ben-Chuan Lin,[1,2,3] Dawei Lu,[1,2,3,4] Yao Lu,[1,2,3] Chudan Qiu,[1,2,3,4] Hui Wang,[5,6,7] Tao Xin,[1,2,3] Shi Yu,[1,2,3] Man-Hong Yung,[1,2,3] Junkai Zeng,[1,2,3] Song Zhang,[1,2,3] Youpeng Zhong,[1,2,3] Xinhua Peng,[6,7,8] Franco Nori,[9,10] and Dapeng Yu[1,2,3,4,*]

[1] *Shenzhen Institute for Quantum Science and Engineering, Southern University of Science and Technology, Shenzhen 518055, China.*
[2] *International Quantum Academy, Shenzhen 518048, China.*
[3] *Guangdong Provincial Key Laboratory of Quantum Science and Engineering, Southern University of Science and Technology, Shenzhen 518055, China.*
[4] *Department of Physics, Southern University of Science and Technology, Shenzhen 518055, China*
[5] *Hefei National Research Center for Physical Sciences at the Microscale and School of Physical Sciences, University of Science and Technology of China, Hefei 230026, China*
[6] *CAS Center for Excellence in Quantum Information and Quantum Physics, University of Science and Technology of China, Hefei 230026, China*
[7] *Hefei National Laboratory, University of Science and Technology of China, Hefei, 230088, China*
[8] *CAS Key Laboratory of Microscale Magnetic Resonance and School of Physical Sciences, University of Science and Technology of China, Hefei 230026, China*
[9] *Quantum Computing Center and Cluster for Pioneering Research, RIKEN, Wakoshi, Saitama 351-0198, Japan*
[10] *Physics Department University of Michigan, Ann Arbor, Michigan 48109-1040, USA*

Quantum computers have made extraordinary progress over the past decade, and significant milestones have been achieved along the path of pursuing universal fault-tolerant quantum computers. Quantum advantage, the tipping point heralding the quantum era, has been accomplished along with several waves of breakthroughs. Quantum hardware has become more integrated and architectural compared to its toddler days. The controlling precision of various physical systems is pushed beyond the fault-tolerant threshold. Meanwhile, quantum computation research has established a new norm by embracing industrialization and commercialization. The joint power of governments, private investors, and tech companies has significantly shaped a new vibrant environment that accelerates the development of this field, now at the beginning of the noisy intermediate-scale quantum era. Here, we first discuss the progress achieved in the field of quantum computation by reviewing the most important algorithms and advances in the most promising technical routes, and then summarizing the next-stage challenges. Furthermore, we illustrate our confidence that solid foundations have been built for the fault-tolerant quantum computer and our optimism that the emergence of quantum killer applications essential for human society shall happen in the future.

## CONTENTS

## I. INTRODUCTION

Quantum computing exploits phenomena of quantum nature, such as superposition, interference, and entanglement, to provide beyond-classical computational resources. Its ultimate goal is to build a quantum computer that can be significantly more powerful than classical computers in solving certain tasks. Historically, quantum computing dates back to the early 1980s, when Benioff developed a quantum-mechanical model of the Turing machine [1], and Feynman [2] and Manin [3] proposed the idea of harnessing the laws of quantum mechanics to simulate phenomena that a classical computer could not feasibly do. In 1994, Shor devised an effi-cient quantum algorithm for finding the prime factors of an integer, a very concrete and important problem for which no efficient classical algorithm is known [4]. Shor's algorithm, along with a number of other quantum algorithms [5], strengthened the foundations of quantum computing, inspired the community of quantum physicists and stimulated research in finding actual realizations of quantum computing. The first implementation scheme came in 1995, when Cirac and Zoller made a proposal for quantum logic gates with trapped ions [6]. In the following years, other physical routes to realize quantum computing, such as nuclear magnetic resonance (NMR) [7–9], spin qubits [10, 11] and superconducting qubits [12], were proposed, and there has been substantial experimental progress in the area since then. Several hardware platforms, including cavity quantum electrodynamics systems, ion traps, and NMR, have successfully realized more than one qubit in experiments since the start of this century. In the following decade, various platforms have achieved quantum information processing on small-scale quantum systems composed of several qubits. In recent years, the field has advanced to the point where research groups have been able to demonstrate quantum devices at a scale around or even beyond forty qubits, particularly in trapped ions and superconducting circuits.

* yudp@sustech.edu.cn

Remarkable progress has been achieved toward fault-tolerant quantum computing. In the beginning, as Fig. 1 shows, universal quantum gates and precise readout have been realized in various physical qubit systems and demonstrated the fulfillment of DiVincenzo criteria. Hardware-level developments and the progress in fabrication further enable the integration of qubits. These achievements enable excessive trials of prototype demonstration of quantum computing, including analog/digital quantum simulation, quantum error correction (QEC), fault-tolerant quantum operations, quantum algorithms, etc. Google achieved quantum supremacy using randomized circuit sampling on their 53-qubit sycamore processor [13] . Afterward, several "quantum advantage" experiments, including superconducting systems [14, 15] and photons [16–18] have been realized, and the gap between the computational power of quantum computers and their classical counterparts was greatly widened. Another milestone is that the realization of quantum annealing in commercialized quantum machines triggers the industrialization of quantum computing. Efforts have been made to develop specialized quantum computers for certain tasks, such as the D-wave annealing machines [19] and the aforementioned photonic boson sampling circuits [16–18]. Moreover, practical QEC has been explored in various physical systems, such as superconducting circuits [20–24], ion traps, semiconductors [25, 26], and nitrogen-vacancy (NV) centers in diamond [27–29]. Considering its speeding-up strides, the breakthrough toward universal fault-tolerant quantum computation is closely tangible. Another noteworthy achievement is the construction of functional quantum simulators [30, 31], digitally or analogously, towards practical problems in quantum chemistry [32] and condensed matter physics [33].

With ever-increasing abilities to precisely manipulate quantum-mechanical systems, the quantum computing community has been shifting the focus from laboratory curiosities to technical realities, from investigating the underlying physics to solving the engineering problems in building a scalable system, from searching for a well-behaved qubit to seriously addressing the question of how to make our near-term quantum hardware practically useful. During the first decade of the 21st century, superconducting qubits, the leading candidate for building scalable quantum computers, were used to demonstrate prototype algorithms with no more than two programmable qubits in most cases. Many efforts have been spent on proof-of-principle tests of various hardware modules. In 2014, two-qubit gate fidelities, an overall performance metric that evaluates the degree of control of a quantum processor, greater than 99% were achieved for the first time in a multi-qubit superconducting circuit, surpassing the error-correction threshold [34]. Since then, the community has seen a trend of growing system size, with 50–100 qubits integrated into state-of-the-art processors. It is remarkable that the average fidelities across these processors are also advancing; In Google's 53-qubit processor, an average of 99.4% two-qubit gate fidelity was achieved with simultaneous operations across the chip [13]; Such enhanced reproducibility indicates immense engineering efforts in all aspects of the experiment, including design, fabrication, wiring, electronics, and software.

Along this grand trend, we have already entered the second stage of quantum computing—noisy intermediate-scale quantum (NISQ) [35] and cloud service of quantum computers, as shown in the left panel of Fig. 1. NISQ for quantum computing is analogous to the early stage of classical quantum computers, when analog/digital signals are hybrid as the exploration of the limit of information processing and the applications of the computing tasks are limited to several areas. During this stage, logic qubits and operation might reach the break-even point by encoding a limited but enough number of noisy qubits in medium-sized integrated systems. As a result, demonstrations of quantum algorithms can be performed using a small amount of logic qubits. And further quantum advantage utilizing quantum computational algorithms or quantum simulation would also be demonstrated with applications on quantum chemistry, variational quantum computers, quantum machine learning, or quantum optimization. Eventually, it is generally believed that fault-tolerant universal quantum computers would be realized in large-scale and integrated quantum systems.

In addition to the advances in hardware, commercially valuable algorithms and applications are beginning to burgeon [36]. A typical example is the variational quantum eigensolver (VQE) algorithm [37], which is shown to be worthwhile eventually from a two-atom molecule calculation to bigger quantum systems. Algorithms for general purposes, in a similar spirit to the forerunner textbook algorithms—Shor's algorithm and Grover's algorithm [5], were developed recently [38], such as the Harrow-Hassidim-Lloyd (HHL) algorithm [39] and the quantum singular value transformation (QSVT) algorithm [40].

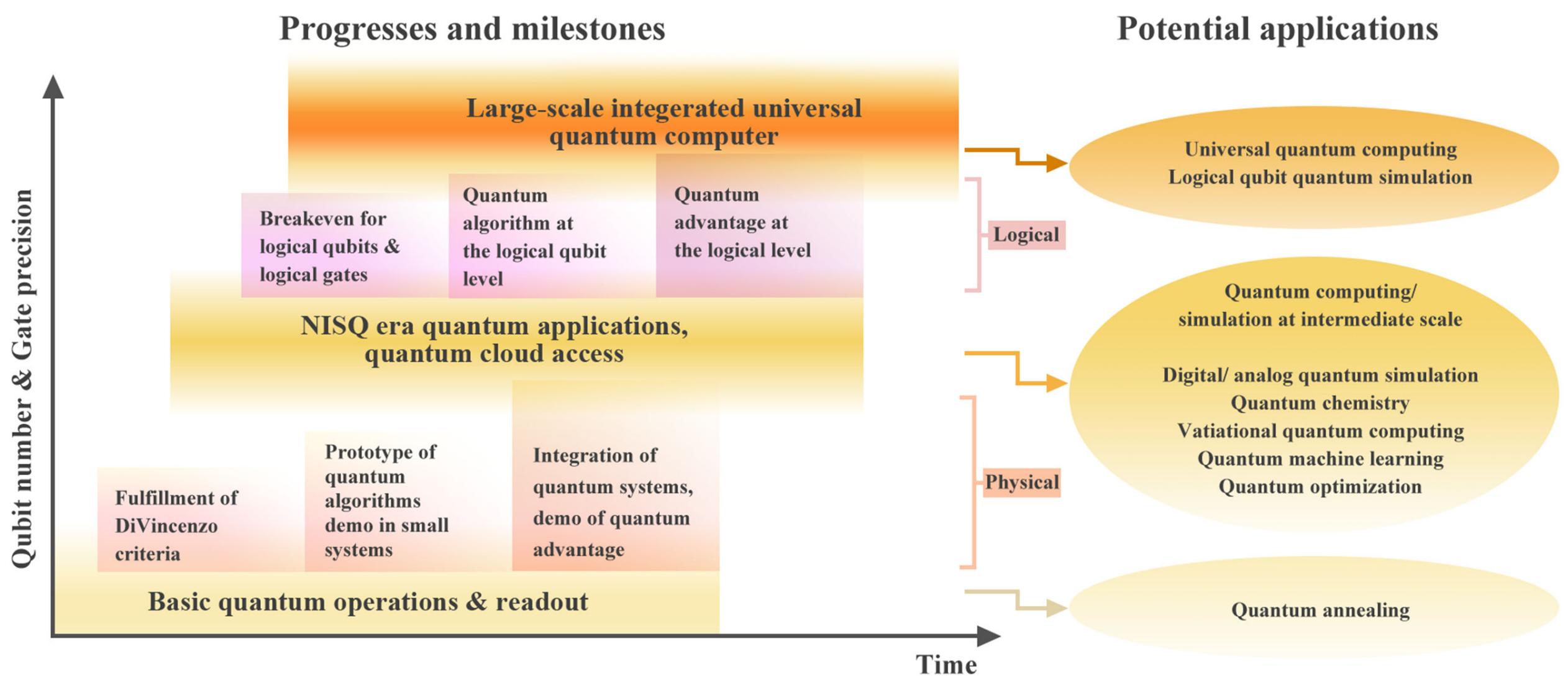

FIG. 1. Quantum computing development levels. The left panel illustrates the three development stages of quantum computing with some iconic progress classified as the physical and logical levels. The right panel lists some potential applications according to different stages. Detailed discussion about this diagram could be referred to the Sec. I.

The paradigm of quantum computing research has been revolutionized over the years from solely academic research. Nowadays, the great impetus comes not only from its intrinsic scientific interest, but also from companies and societies [41]. With the aforementioned tremendous progress, the approach to full-stack quantum computing [41–43] approach is encouraging. As commercialization is becoming a trend, many large companies and start-ups are contributing to this field jointly with the scientific community. We shall see further contributions and incentives to the development of this field coming from commercialization, as has already happened for classical computers, genetic technology, and artificial intelligence. From a broader society scope, cloud quantum computing, such as IBM's quantum network, makes it possible for global users around the world to explore new quantum algorithms without their own hardware devices. As quantum computing lessons and experiences in schools and universities are becoming a routine for the next generation, more well-educated engineers and scientists are being enrolled in the field, equipped with insights and knowledge of quantum science. Thus, the positive feedback from society is creating a new norm for quantum computing research compared to its primitive days.

In this review, we will focus on hardware platforms that have the potential to realize the ultimate large-scale quantum computers, including superconducting circuits, trapped ions, semiconductors, neutral atoms, NMR, NV centers, and photonics. In particular, we will focus more on the important advances in these platforms over the past decade. By following the guidelines of DiVincenzo's criteria [44], we will introduce how to implement the key elements of quantum computing in each physical system and their typical features and advantages. The scalability of each platform and critical challenges in recent developments will also be discussed here. Moreover, recent progress on quantum algorithms will also be mentioned in this review. By combining fast-developing hardware platforms and potential applications, we hope to shed light on the innovations that quantum computing can bring in the foreseeable future.

Topological quantum computation provides another approach to tackling quantum errors by keeping the computational states to the desired pure quantum states without erroneous results. A typical type of topological qubits is made of Majorana zero modes, which are immune to environmental noise and thus overcome the inevitable decoherence at the hardware level through the Majorana non-locality and braiding operations. However, the non-topological in-gap states or trivial zero-energy states can also mimic the typical Majorana behavior, making the detection and other operations of Majorana zero modes difficult. So far, the non-locality and braiding operations to demonstrate the non-Abelian statistics have yet to be proved before the realization of topological qubits. A complete discussion of topological quantum computation is beyond the scope of this review. The interested reader is referred to the literature for further details [45–47].

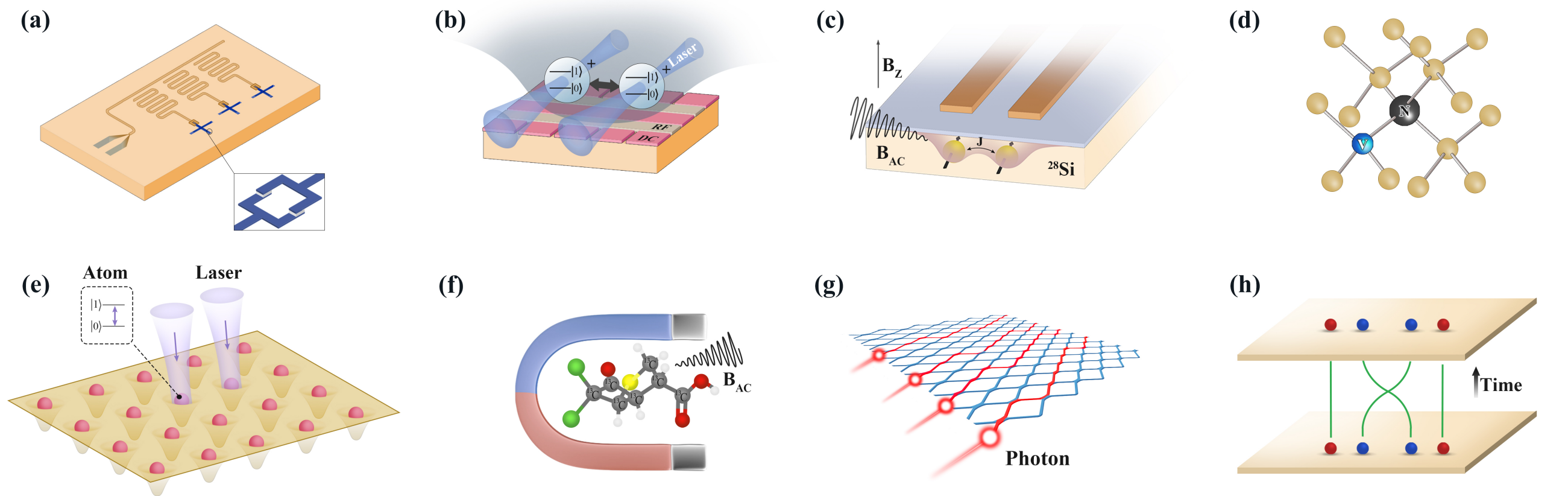

**(a) Superconducting quantum processor:**
An array of artificial atoms made of Josephson junctions and capacitors, fabricated with lithography, controlled with microwave electronics
**Number of qubits with individual control:**
~100
**Gate fidelities:**
Systems with 53 qubits: $F_1 > 99\%$; $F_2 > 99\%$
**Pros:**
Fast gate speed, tailored qubits, high controllability and scalability
**Cons:**
Crosstalk between qubits, low temperature, supporting technology for scaling up

**(b) Trapped ion quantum computing:**
Laser-cooled atomic ions held by radio-frequency electric fields in ultra-high vacuum environment
**Number of qubits with individual control:**
20-30
**Gate fidelities:**
Systems with dozens of ions
$F_1 > 99.9\%$; $F_2 > 99\%$
**Pros:**
Extremely long coherence time and excellent gate operations on few ions;
Reconfigurable connectivity between ion qubits
**Cons:**
Technically challenging in integration

**(c) Semiconductor spin-based quantum computing:**
Electron or hole spins in semiconductor (silicon) quantum dot
**Number of qubits with individual control:** 6
**Gate fidelities:**
Donor spin qubit: $F_1 \sim 99.99\%$; $F_2 \sim 99.5\%$
Gate-defined qubit: $F_1 \sim 99.9\%$; $F_2 \sim 99.51\%$
**Pros:**
Semiconductor fabrication
Long coherence and fast high-fidelity gate
Work at temperature > 1 K
Small footprint
**Cons:**
Challenge of nanoscale fabrication

**(d) NV center quantum computing:**
Point defects in diamond; electron and nuclear spins with long coherence time; atom-like properties and solid-state host environment
**Number of qubits with individual control:**
10 (one electron + nine $^{13}$C nuclear spins)
**Gate fidelities:**
$F_1 \sim 99.995\%$; $F_2 \sim 99.2\%$
**Pros:**
Work at room temperature
Excellent quantum sensor
Handy in quantum network
**Cons:**
Hard to scale up

**(e) Neutral atom array quantum computing:**
Neutral atom arrays trapped in optical tweezers with controlled interactions based on the Rydberg interactions
**Number of qubits in digital quantum processors:** 24
**Number of qubits in analog quantum simulators:** 289
**Pros:**
Both digital and analog quantum simulations
Scaling up beyond 100 qubits in programmable geometries
**Cons:**
Improving fidelities of two-qubit gates

**(f) NMR quantum computing:**
Nuclear spins in molecule
**Number of qubits with individual control:**
12
**Gate fidelities:**
$F_1 \sim 99.98\%$; $F_2 \sim 99.3\%$
**Pros:**
Work at room temperature
Long coherence time
Digital quantum simulator
**Cons:**
Hard to scale up

**(g) Photonic quantum computing:**
Coherently manipulating a large number of single photons or canonically conjugate pair of variables for electromagnetic modes to process quantum information
**Number of qubits with individual control:**
18
**Number of photons with coherent control**:
255 (Jiuzhang 3.0 (to be published))
**Gate fidelities:**
$F_1 \sim 99.84\%$; $F_2 \sim 99.69\%$
**Pros:** robust against decoherence, working at room temperature, compatible with CMOS fabrications, natural interface for distributed quantum computing
**Cons:** Big challenge to realize deterministic photon-photon gate

**(h) Topological quantum computing:**
Fault-tolerant quantum computation based on non-abelian braiding of anyons
**Number of qubits with individual control:**
N/A
**Gate fidelities:**
N/A
**Pros:**
Intrinsic topological protection
Few physical qubits to construct a logic qubit
Promising to achieve large-scale, error-corrected computation
**Cons:**
The ideal materials or systems not found yet.
Zero topological qubit so far

FIG. 2. Schematic summary of different types of quantum bits (top half) and their corresponding pros and cons. (bottom half). $F_1$ ($F_2$) is the one-qubit (two-qubit) gate fidelity.

## II. QUANTUM ALGORITHMS

*Introduction.—* It is anticipated that quantum computers utilizing the exotic quantum features can solve computational problems more efficiently than their classical counterparts. For example, in the query model, given an oracle access to a function $f : \{0,1\}^n \to \{0,1\}$, a classical computer can only query it once at a time, whereas a quantum computer can query the oracle once and obtain all $2^n$ values simultaneously, a phenomenon known as quantum parallelism. Formally,

$$\sum_x |x\rangle |0\rangle \mapsto \sum_x |x\rangle |f(x)\rangle \qquad (1)$$

can be achieved on a quantum computer. However, quantum parallelism alone is not useful because when one performs a measurement, the quantum state collapses, and only one bit of information can be obtained. To design quantum algorithms, quantum parallelism needs to be combined with other features such as interference and entanglement.

In 1985, Deutsch combined quantum parallelism with interference to design the first quantum algorithm that can solve a black-box problem with fewer queries than a classical computer [48]. Specifically, in Deutsch's problem, one is given a function $f : \{0,1\} \to \{0,1\}$ and asked whether the function is constant, that is, $f(0) = f(1)$, or not. Classically, we would need two queries to solve this problem; but with quantum computers, only one query is needed. Later, it was generalized to a multi-qubit version called the Deutsch-Jozsa algorithm, which can achieve an exponential speedup over any classical deterministic algorithms [49]. However, the quantum speed-up vanishes in the presence of a small error probability. In 1993, Berstein and Vazirani proposed another problem and designed a quantum algorithm for it that can achieve polynomial speedup even over classical randomized algorithms [50]. After one year, Simon strengthened their result by designing Simon's problem and a quantum algorithm for it that yields an exponential speedup [51].

These early-stage explorations focused mostly on the search for problems that quantum computers can solve more efficiently than classical computers, instead of focusing on real-world applications. But interestingly, as it turned out later, Simon's algorithm inspired Shor to design quantum algorithms to solve discrete logarithmic and integer factoring problems [4], which are widely used in cryptography.

*Quantum Fourier transform.*— In the next stage of the development of quantum algorithms, several quantum algorithmic primitives emerged and appeared to be extensively used in designing new quantum algorithms. One such primitive is the quantum Fourier transform (QFT), which implements the Fourier transform matrix

$$(F_N)_{jk} := \omega_N^{jk}/\sqrt{N} \tag{2}$$

with a polynomial-sized quantum circuit on a quantum computer, where $N := 2^n$ and $\omega_N := e^{2\pi i/N}$ is the $N$-th root of unity. Simon's algorithm uses a special instance of QFT, namely the Hadamard transform, which corresponds to the case $N = 2$ and $\omega_N = -1$. From a group-theoretic point of view, $F_N$ is the Fourier transform over $\mathbb{Z}_N$, the additive group of integers modulo $N$, consisting of elements $\{0, 1, \cdots, N-1\}$; Simon's Hadamard transform is the Fourier transform over $\mathbb{Z}_2^n$ [52].

There are two steps in Shor's factoring algorithm, a classical polynomial-time reduction from integer factoring to period finding, followed by an efficient quantum algorithm for solving the period finding problem [4], which uses QFT over $\mathbb{Z}_N$. Combining these two steps, Shor obtained a polynomial-time quantum algorithm for solving integer factorization, which has super-polynomial speedup over the best classical algorithm. Kitaev gave a generalized QFT over an arbitrary finite Abelian group, with which he designed a polynomial-time quantum algorithm for finding the stabilizer of an Abelian group; the Abelian stabilizer problem includes integer factoring and discrete logarithm as special instances [53]. It is worth mentioning that Kitaev also gave the phase estimation algorithm in the same paper, which estimates the phase $\phi$ in $U|\psi\rangle = e^{i2\pi\phi}|\psi\rangle$ and can be used to solve the period finding problem [53]. In a coherent picture, all these problems belong to the hidden subgroup problems category [52, 54].

*Quantum search.*— The second primitive starts from Grover's search algorithm [55, 56], which concerns searching over an unsorted database for a target. Formally, given a function $f : \{0,1\}^n \to \{0,1\}$ and the promise that there is exactly one $x_0$ such that $f(x_0) = 1$, the search problem is to find the target $x_0$. Since there is no structure in this problem, a classical algorithm will need $\Omega(2^n)$ times of queries to find the target $x_0$ with sizable probability. Grover's algorithm allows a quantum computer to find the target with $O(\sqrt{2^n})$ queries to the database, which achieves a quadratic speedup over classical computation. Grover's algorithm repeatedly applies the Grover iterate

$$G = (2|u\rangle\langle u| - I)(I - 2|x_0\rangle\langle x_0|) , \tag{3}$$

which is the product of two reflections; here, $|u\rangle := \frac{1}{\sqrt{N}}\sum_y |y\rangle$ is the uniform superposition and $I - 2|x_0\rangle\langle x_0|$ is the quantum query operator. Grover's algorithm is optimal in the sense that any quantum algorithm that solves this problem will take at least $\Omega(\sqrt{2^n})$ queries [57]. Grover's algorithm can be extended to amplitude amplification, which can handle the case of multiple numbers of targets [58–60]. More precisely, given a quantum (or classical) algorithm $\mathcal{A}$ applied to $|0^n\rangle$ that can output a correct target when measured with probability $p$, it would require running the algorithm $1/p$ times to obtain the targeted result. But amplitude amplification can obtain the target in time $O(1/\sqrt{p})$, which is also a quadratic speedup. The fixed-point version of Grover's algorithm or amplitude amplification can even handle the case when the number of targets is unknown [61–63]. Grover's algorithm has inspired more applications than Shor's algorithm, as it can be used to speed up the search subroutine for solving many optimization problems [64–68].

One may consider the search problem in an alternative paradigm, namely, the Markov chains or random walks. The quantum version of random walks includes the continuous-time quantum walk [69–71] and the discrete-time quantum walk; we focus on the latter here. The framework of the discrete-time quantum walk was incrementally developed in several works [72–78]. Later, this framework was applied to obtain a different formulation of Grover's search algorithm [79]. In a breakthrough work, Ambainis designed a quantum walk algorithm for element distinctness [80] that achieves better query complexity than a direct application of Grover's algorithm and matches the theoretical lower bound [81]. Ambainis' result was generalized subsequently [82, 83], and, in particular, Szegedy gave a general framework for quantizing classical Markov chains [83], which was further improved in [84]. This quantum walk-based search algorithm finds many applications [85, 86], including triangle finding [87], testing group commutativity [88], etc.

*Hamiltonian simulation.*— The third primitive that will be discussed in this review is Hamiltonian simulation, which approximates the time evolution operator $e^{-iHt}$ of a Hamiltonian $H$ on a quantum computer. In fact, Hamiltonian simulation is one of the initial motivations for developing quantum computing [2]. The first quantum algorithm for implementing the time evolution operator is given by Lloyd [89], which is based on the Lie-Trotter formula. For example, suppose that we have a local Hamiltonian $H = H_1 + H_2$ such that the time evolution of the local terms $H_1$ and $H_2$ can be efficiently implemented on a quantum computer. The Lie-Trotter formula gives

$$e^{-iHt} = (e^{-iH_1t/s}e^{-iH_2t/s})^s + O(t^2/s) , \quad (4)$$

which means that $e^{-iHt}$ can be implemented by alternating $H_1$ and $H_2$ over an incremental time $t/s$. One can also use the higher-order formula [90, 91] to approximate the time evolution of $H$. The general scheme is called the product-formula approach, or Trotterization, which was later applied to simulate a sparse Hamiltonian [92–94]. Later, a Hamiltonian simulation method based on quantum walk [95, 96] was proposed, which achieved linear gate and query complexities in the evolution time $t$, matching the lower bound imposed by no fast-forwarding theorem [93].

Another important approach to simulate Hamiltonian dynamics is using a linear combination of unitaries (LCU) [97], and it is shown to have the optimal dependence on the simulation precision [98, 99]. The LCU approach is combined with the quantum walk approach to give an algorithm that has optimal dependence on all parameters of interest, such as precision, sparsity of the Hamiltonian, the evolution time, etc. [100]. Moreover, there is a subroutine used in the LCU approach that was later named block encoding, which turns out to provide a versatile toolkit in designing quantum algorithms. In a series of works, Low et al. gave improved Hamiltonian simulation algorithms based on block encoding and quantum signal processing [101–104]. The idea is to encode the Hamiltonian as a block of a unitary, and then apply the polynomial transformation to the Hamiltonian using the quantum signal processing technique [101]. This method was further generalized to a framework called QSVT [105], which covers most existing quantum algorithms as special cases, achieving a grand unification of quantum algorithms [106]. However recently, an in-depth analysis of the Trotter error showed that the product-formula approach can achieve a competitive scaling of gate complexity compared to other approaches [107].

*Quantum linear algebra and quantum machine learning.*— The previous primitives can be combined to design new quantum algorithms. Here, we discuss quantum algorithms for linear algebra and machine learning. Quantum linear algebra starts with the HHL algorithm, named after Harrow, Hassidim, and Lloyd [39]. The problem they considered is to solve linear systems of equations; that is, given a matrix $A$ and a vector $\mathbf{b}$, solve $A\mathbf{x} = \mathbf{b}$ for $\mathbf{x}$. Given a quantum state $|b\rangle$ that encodes the vector $\mathbf{b}$ in its amplitudes, HHL uses Hamiltonian evolution and phase estimation to approximately prepare the state $|x\rangle = A^{-1}|b\rangle$. Provided that the whole description of the solution $\mathbf{x}$ is not required and that the state $|b\rangle$ can be prepared, the HHL algorithm can achieve exponential speedup over any classical algorithm [39]. HHL was applied to many quantum machine learning algorithms to obtain exponential quantum speedup, including quantum $k$-means clustering [108], quantum principal component analysis [109], quantum support vector machine [110], quantum data fitting [111], etc; see Ref. [112] for a review of these algorithms.

However, it is not clear whether such an exponential quantum speedup is artificial or not. Specifically, these quantum machine learning algorithms typically made strong input assumptions, such as quantum random access memory (QRAM) with access to the classical data [113]. It might be possible to derive efficient classical algorithms in an analogous setting. In 2018, the breakthrough work by Tang [114] gave a classical algorithm that dequantizes the quantum algorithm for recommendation systems [115], which was previously believed to have an exponential speedup, with only a polynomial slowdown. Tang's result stimulated a series of subsequent work on dequantizing various quantum machine learning algorithms, such as those for principle component analysis [116], solving low-rank linear systems [117, 118], solving low-rank semidefinite programming [119], etc. The sample and query access model [116] to the input data is assumed in those works, which is a classical analogue to the input assumptions in many quantum machine learning algorithms. Since the QSVT provides a primitive for unifying quantum algorithms, especially quantum linear algebra, these dequantization results were later extended to a unifying framework by dequantizing the QSVT [40]. Therefore, whether exponential quantum speedup can be achieved in machine learning is still under debate.

*Variational quantum algorithms.*— Apart from quantizing machine learning algorithms with HHL, another exploration is inspired by neural networks, which are variational quantum algorithms. Variational quantum algorithms are hybrid quantum algorithms that prepare parameterized quantum states on a quantum computer and use classical computers to optimize the parameters. The first variational quantum algorithm is the VQE [37], designed to tackle quantum chemistry problems. Its goal is to find the ground state and ground energy of local Hamiltonians. Before VQE, a common approach was to use quantum phase estimation [120, 121]. However, such an approach, just like other quantum algorithms, imposes a stringent coherence requirement on the quantum devices, which is challenging in the current NISQ era [35].

Since VQE, more works have been done in this direction. Inspired by quantum adiabatic algorithm [122], Farhi et al. proposed the quantum approximate optimization algorithm (QAOA) for solving combinatorial optimization problems such as the max-cut problem [123]. The third family of variational quantum algorithms is the quantum neural networks, which aims to solve machine learning problems such as classification [124, 125] and generative modeling [126, 127].

In the current NISQ era, although we have demonstrated quantum computational supremacy in various models [13, 15, 128, 129], these problems are not designed to be of practical relevance. Variational quan-tum algorithms are regarded as promising approaches for demonstrating "killer applications" on quantum computers. Such applications might appear in various areas including quantum chemistry, material science and biological science. For example, in quantum chemistry, VQE can be used to compute the low-energy eigenstates of electronic Hamiltonians, which helps understand chemical reactions and design new catalysts [130]. As for biological science, optimization is often involved in many fields like sequence analysis and functional genomics [131]. This opens opportunities for potential quantum speedup by using quantum neural networks, and quantum variational auto-encoders [132], etc. However, there is a long way to go along this path and continuous efforts should be put into the study of variational quantum algorithms. Moreover, to make it practical for the neat-term quantum devices, perhaps error mitigation techniques are also required [133–137].

## III. SUPERCONDUCTING QUBITS

*Introduction.*— Superconducting qubits are nonlinear superconducting circuits based on Josephson junctions, with quantized electromagnetic fields in the microwave frequency domain (typically 0.1–12 GHz). They operate at cryogenic temperatures ($\sim$ 10 mK; equivalent to $k_{\mathrm{B}}T/h \sim 0.2$ GHz) provided by dilution refrigerators in order to suppress thermal fluctuations. Superconducting qubits recently emerge as a leading platform for scalable quantum information processing. Some recent milestones include the demonstration of quantum supremacy using a 53-qubit superconducting quantum processor [13], which is further strengthened with a 66-qubit processor [128]. Offering scalable high-fidelity control and configurable interactions, superconducting circuits have become a versatile playground for quantum computational tasks [125, 128, 138–141], quantum simulation [142–150], quantum annealing [19, 151], quantum chemistry [152–155], exotic many-body physics [156–161], new regimes for light-matter interaction [162–165], quantum sensing [166, 167] and studying biological processes [168].

Some facts about superconducting qubits are summarized in Fig. 2(a), and a list of excellent reviews on superconducting qubits can be found in Refs. [169, 173–190]. The charge carriers in superconductors, known as Cooper pairs, can flow without dissipation, a desirable feature for preserving quantum coherence of a macroscopic system. More importantly, non-trivial quantum properties emerge from the integration of a special superconducting circuit element, the Josephson junction, which is usually in the form of a sandwich structure consisting of two superconducting electrodes separated by a nanometer-thick insulating layer (Fig. 3a); Cooper pairs can tunnel through the insulating barrier with a supercurrent no larger than the critical current $I_{\mathrm{c}}$ of the junction which depends on the material, thickness, and size [191, 192]. From a circuit point of view, a Josephson junction can be modeled as a native capacitor $C_{\mathrm{J}}$ in parallel with a nonlinear inductor $L_{\mathrm{J}} = \Phi_0/2\pi I_{\mathrm{c}} \cos\phi$, where $\Phi_0 = h/2e$ is the superconducting flux quantum and $\phi$ is the superconducting phase difference across the junction. Two characteristic parameters of a Josephson junction are its Josephson energy $E_{\mathrm{J}} = \Phi_0 I_{\mathrm{c}}/2\pi$ and the charging energy $E_{\mathrm{C}} = e^2/2C_{\mathrm{J}}$.

*Qubit construction.*—There have been numerous explorations of how to construct a superconducting qubit using Josephson junctions. Traditionally, superconducting qubit designs are categorized into charge [193], flux [194, 195] and phase qubits [196]; all are successful in many early demonstrations [12, 197–208]. In recent years, the transmon qubit [209] and a modified version for implementing scalability, i.e., the Xmon qubit [171] have become popular. These modified designs of the charge qubit shunt a Josephson junction with a large capacitor $C_{\mathrm{S}}$ to strongly suppress their sensitivity to the charge fluctutaions [210]. Typically, this shunt capacitor lowers the effective charging energy $E_{\mathrm{C}} = e^2/2(C_{\mathrm{S}} + C_{\mathrm{J}})$ to the regime of $E_{\mathrm{J}}/E_{\mathrm{C}} > 50$ (Fig. 3b); as a result, the sensitivity to charge fluctuations is strongly suppressed. The fact that the transmon design has the simplest possible circuit geometry makes it more tolerant of fabrication variations and excellent at reproducibility. The Hamiltonian of the transmon qubit is the same one for the charge qubit and can be expressed as

$$H = 4E_{\mathrm{C}}n^2 - E_{\mathrm{J}}\cos\phi, \tag{5}$$

where $n$ is the number of Cooper pairs that traverse the junction; $n$ and $\phi$ satisfy the commutation relation $[\phi, n] = i$. Note that the Hamiltonian is identical to the one that describes a quantum particle in a one-dimensional potential (Fig. 3c). In the $C_S \gg C_J$ limit, the low-energy eigenstates are, to a good approximation, localized states in the potential well, and the superconducting phase $\phi$ is small. We can therefore expand the potential term into a power series:

$$-E_{\mathrm{J}}\cos\phi = \frac{1}{2}E_{\mathrm{J}}\phi^2 - \frac{1}{24}E_{\mathrm{J}}\phi^4 + \mathcal{O}(\phi^6). \tag{6}$$

The first quadratic potential term leads to a quantum harmonic oscillator with equidistant energy levels $\hbar\omega_{10}$, whereas the second quartic potential arising from the Josephson nonlinearity introduces anharmonicity to the level structure, allowing the transition energy $\hbar\omega_{21}$ between first-excited state $|1\rangle$ and second excited state $|2\rangle$ to be different from that between the ground state $|0\rangle$ and first-excited state $|1\rangle$. This nonlinearity allows one to define a qubit in the computational subspace consisting of the lowest two energy levels $|0\rangle$ and $|1\rangle$ only. The design may be further modified by replacing a single junction with a pair of junctions so that the effective Josephson energy, and consequently the qubit frequency, can be tuned by adjusting the magnetic flux threading the two-junction loop.

The transmon design can be implemented with a qubit circuit embedded in a three dimensional cavity (Fig. 3d)

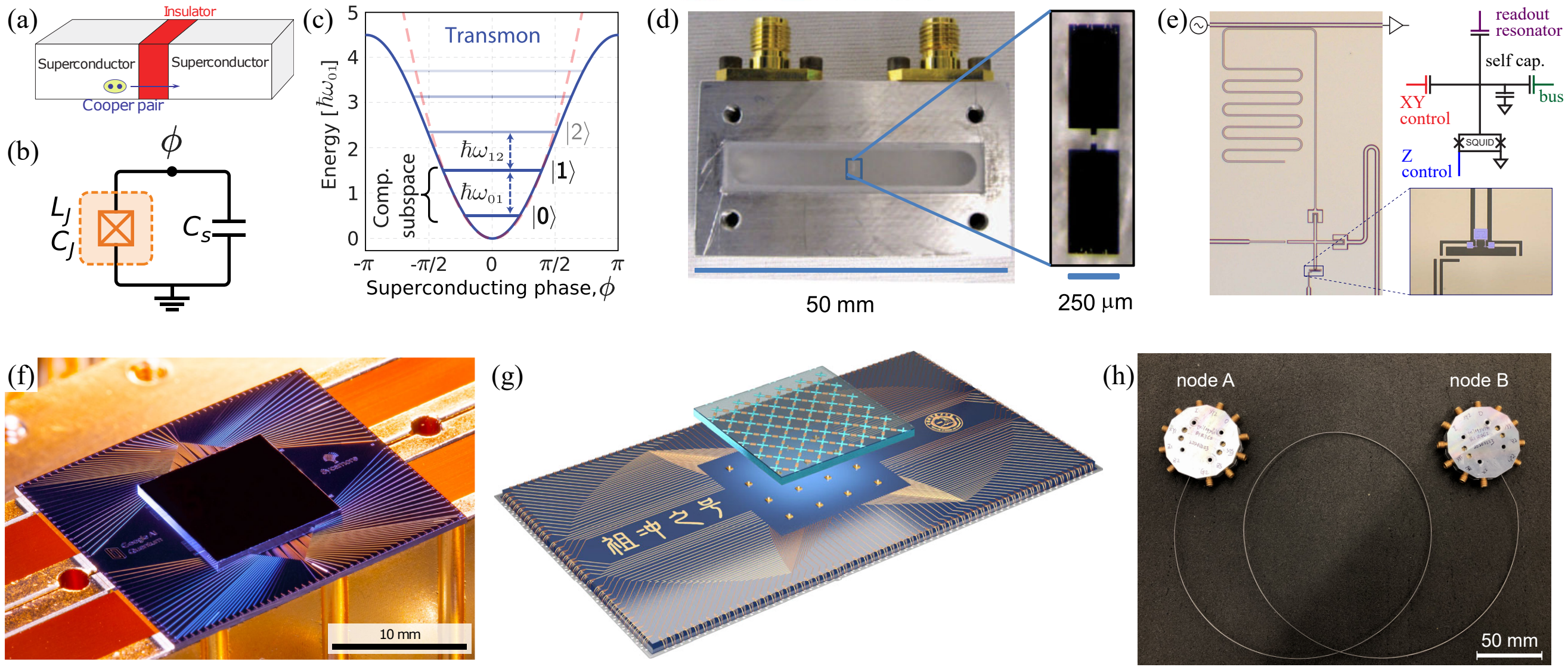

FIG. 3. (a) Schematic of a Josephson junction composed of two superconductors separated by a thin insulating layer through which Cooper pairs can tunnel. Adapted from Ref. [169], Springer Nature Limited. (b) Circuit diagram of a transmon qubit consisting of a Josephson junction (Josephson inductance $L_{\mathrm{J}}$, self-capacitance $C_{\mathrm{J}}$) and a shunt capacitor $C_{\mathrm{S}}$ ($C_{\mathrm{S}} \gg C_{\mathrm{J}}$). (c) Potential profile and level diagram of the transmon qubit, a quantum anharmonic oscillator. (d) Image of a transmon qubit embedded in a three dimensional cavity. Adapted from Ref. [170]. (e) Image of a planar transmon qubit. Adapted from Ref. [171]. (f) Photograph of the Sycamore quantum processor. Adapted from Ref. [13], Springer Nature Limited. (g) Device schematic of the *Zuchongzhi* quantum processor. Adapted from Ref. [128]. (h) Photograph of a modular quantum processor consisting of two nodes. Adapted from Ref. [172], Springer Nature Limited.

and in the form of lithographically defined circuits based on superconducting materials such as aluminum and niobium (Fig. 3e) and many other variants [170, 171, 211–213]. Qubit designs with alternative topology such as capacitively shunted flux qubit [210, 214–217], fluxonium [218–221], and 0-$\pi$ qubit [222] have also been under active development and shown promising progresses. By engineering the energy-level spectra and the coupling matrix elements, some of these designs have a better-defined two-level system and intrinsic protection against external perturbations at the cost of increased circuit complexity. The remarkable flexibility in configuring the Hamiltonian offers a rich parameter space to search for desired qubit properties and therefore gives superconducting qubits the name "artificial atoms".

*Readout and initialization.*—Having a well-defined two-level system is not enough for quantum computing; the ability to faithfully measure and initialize the qubit is also indispensable. The prevailing technique for discerning the qubit state is the dispersive readout scheme. Utilizing the cavity or circuit quantum electrodynamics (cQED) architecture, a qubit is strongly coupled to but sufficiently detuned from a readout resonator [223, 224]; the qubit induces a state-dependent shift in the resonator frequency from which the qubit state can be inferred by interrogating the resonator. The cQED scheme has been successful in achieving fast, high-fidelity, non-demolition readout, assisted by a list of technologies that have been invented around this approach. To avoid extra decoherence introduced by the readout resonator, a Purcell filter can be placed between the resonator and the external circuitry to reshape the environmental mode density seen by the qubit and the resonator [225–228]; in this way, one may enhance the readout speed while inhibiting qubit relaxation. In addition, the use of Josephson parametric amplifiers (JPA) [229–237] at the beginning stage of the readout signal amplification can also bring an immediate improvement to the measurement fidelity. It is noteworthy that other techniques including multiplexed readout [238], multilevel encoding [239], and photon counting method [240] also help improve the measurement efficiency and scalability. Between consecutive measurements, superconducting qubits are typically initialized by simply waiting for the qubit to relax to its ground state. Various conditional and unconditional reset techniques have been developed for superconducting qubits to accelerate this process [241–245].

*Gates.*—Controlling superconducting qubits is challenging because performing a quantum logic or unitary operation is fundamentally an analog process governed by the Schrodinger equation and the realistic Hamiltonian is often far from ideal. A single qubit XY operation, rotation around an axis in the XY plane in the Bloch sphere, is commonly implemented by driving the qubit with a resonant microwave pulse. For weakly anharmonic qubits such as the transmon, the resonant drive can induce unwanted leakage to higher excited states and additional phase errors; the derivative-removal-by-adiabatic-gate (DRAG) scheme, which adds additional quadrature components to the pulse, has become a routine in pulse calibration to combat these coherent errors at no additional hardware cost [246, 247]. The single-qubit phase gate or Z gate, rotation around the Z axis, can be realized either by combining XY rotations or by adding a physical Z pulse provided that the qubit frequency is adjustable or by performing the more efficient virtual Z gate through shifting the phases of XY rotations [248]. Heat dissipation is another important concern for cryogenic experiments when scaling to a large number of qubits; a more energy-efficient approach for single-qubit operations using single-flux-quantum (SFQ) circuits has been demonstrated recently [249].

Entangling operations are currently the performance bottleneck for existing quantum processors. Among the numerous entangling gate schemes, most of them are between two qubits and they generally belong to two families. One general approach is to frequency-tune the relevant energy levels into resonance to initiate interactions; related demonstrations include the implementation of iSWAP or $\sqrt{\text{iSWAP}}$ gate [250, 251] and the controlled-Z gate [252–257]. The other method is to apply microwave pulses at certain non-local transitions; for example, the cross-resonance gate [258, 259], resonator-induced phase gate [260] and parametrically driven gate [261, 262]. The gist for obtaining high-fidelity two-qubit gates is to engineer an effective interaction strength such that it is strong during gate operation for achieving short gate time while as weak as possible outside of the gate window for avoiding unwanted interactions, in another word, a high on/off ratio. In a fixed-coupling architecture where the qubit-qubit coupling strength $g$ is almost constant, a straightforward way to turn the interaction on or off is to tune the qubit frequencies into or away from resonance. However, as the qubit-qubit connectivity increases, each qubit sees more transitions in its surroundings; it becomes increasingly difficult to manipulate the whole system in a clean fashion. This is known as the frequency-crowding problem, one of the main challenges in scaling up quantum processors. The problem also exists for alternative coupling schemes such as the all-to-all connection via a bus resonator [263]. It may be resolved by the tunable-coupling schemes in which the coupling strength $g$ can be independently controlled over a large dynamic range [264–271]. In recent years, a tunable-coupling architecture based on native capacitive coupling and interference effect [272] has become a trending solution; many research groups have made tremendous progress in gate fidelities, including some results approaching the 99.9% mark [273–280]. Typical performance of superconducting processors is summarized in Table I.

*Decoherence.*—Quantum information can be quickly destroyed by decoherence. The bad news is that superconducting qubits are extremely susceptible to external fluctuations due to their macroscopic nature. One immediate solution is, of course, to make the qubit lifetime longer. Ever since the first observation of quantum coherence in superconducting qubits [12], the lifetime of the qubit has been improved by six orders of magnitude from nanoseconds to milliseconds [282–284] in about 20 years. This remarkable progress is attributed to a combination of advances in design, material, fabrication quality, and testing environment. The current common belief is that spurious two-level systems (TLS) that reside in the vicinity of the qubits are a major source of decoherence [285] and unpredictable fluctuations in coherence and qubit frequencies, which can be troublesome in large-scale implementations [286–290].

Besides coherence improvement on the hardware side, another way to combat noise and decoherence is through quantum control methods. A particularly useful technique is dynamical decoupling (DD) [291] which uses tailored pulse sequences to correct for coherent noise, in particular, the notorious $1/f$ noise [292] that is ubiquitous in these solid-state devices. Designing an optimal sequence requires detailed knowledge of the noise such as its spectral properties which may be extracted using various techniques at different frequency ranges [251, 293–298].

Given the limited coherence, the performance of a quantum processor may also be enhanced through optimized quantum compiling, i.e., to translate high-level operations into a shorter sequence of native gates [299]. Compiling on a superconducting quantum processor can be exceptionally challenging due to the planar geometry and limited connectivity; often, the final sequence to execute is too time-consuming. An effective strategy is to fully explore the hardware capabilities and diversify the available gate alphabets to optimize compilation. Recent progress on continuous gate set, multi-qubit gates, and qudit operations have shown considerable potential in this respect [275, 300–304].

*Quantum error correction.*—Since the state-of-the-art gate error rate ($10^{-3}$) is many orders of magnitude higher than that a logical qubit would require ($10^{-12}$–$10^{-15}$), QEC is necessary for building a universal quantum computer. Surface codes [305, 306], which encode logical qubits into a square lattice of physical qubits, are appealing for planar architectures. Recently, we have observed a surge of exciting experimental developments in this respect [21, 22, 24, 307–309]. In some of these experiments the performance is getting close to or partially exceeds the QEC threshold. Still, it is challenging to achieve substantial error-correction gain and most importantly, to have the performance reproducible at an even larger scale. In the near future, a logical qubit made of a few hundreds to a thousand physical qubits is highly anticipated; in the next five to ten years, we may have an idea about whether a fault-tolerant quantum computer is feasible and how powerful it can be. A relevant issue with increasing attention is the presence of cosmic rays which can cause chip-wide failure and is catastrophic to surface codes [310–312].

Another promising route to QEC is bosonic codes, where logical qubits are encoded in microwave photon

TABLE I. Typical performance reported in superconducting qubits.

| No. of qubits | $T_1$ ($\mu$s) | $T_2^*$ ($\mu$s ) | $t_{1q}$ (ns) | $Err_{1q}(10^{-3})$ | $t_{2q}$ (ns) | $Err_{2q}(10^{-3})$ | $t_r$ ($\mu$s ) | $Err_r$ ($10^{-2}$) | Fridge | Size |
|---|---|---|---|---|---|---|---|---|---|---|
| 53, 66 [13, 128] | $16 - 30.6$ | $\sim 5.3$ | $\sim 25$ | $\sim 1.4$ | $\sim 12$ | $\sim 5$ | $\sim 1$ [281] | $\sim 3.1$ [281] | $\sim 20$ mK | $\sim 1$ mm$^2$ |
| <10 [273–280] | $15 - 76$ | $12 - 105$ | $\sim 30$ | $\sim 1$ | $10 - 200$ | $\sim 1.5$ | | | | |

$t_{1q}$ ($t_{2q}$) is the duration of one-qubit (two-qubit) gate.
$t_r$ ($Err_r$) stands for the duration (error rate) of the readout.

states of three-dimensional superconducting cavities. Depending on how a logical qubit is encoded into harmonic oscillator states, there are different kinds of bosonic codes [313–315]. The cat codes and associated variants utilize a superposition of two photonic cats of the same parity as logical qubit [316–320]; the binomial codes instead use definite photon number parity as code words [321, 322]; the Gottesman-Kitaev-Preskill (GKP) codes implement a coherent state lattice in phase space [323–327] with the advantage that errors, measurements, and gates are simple displacements of the oscillators. To date, only bosonic codes have reached the break-even point in QEC experiments, which means that the error-corrected qubit has longer lifetime than the otherwise. This is due to the fact that microwave photons have fewer error syndromes and three-dimensional cavities usually have higher quality.

*Scalability.*—Lastly, we would like to touch upon the most concerning question: how to make the superconducting quantum processor more scalable. With the continuous improvement in planar circuit design and fabrication and the development of flip-chip packaging, dozens of superconducting qubits have been integrated on a single processor so far, allowing for the demonstration of quantum supremacy [13] (Fig. 3f) or quantum computational advantage [128] (Fig. 3g). It is worth emphasizing that simply printing thousands of qubits is straightforward, but the real challenge is to achieve high-fidelity operations for all qubits, simultaneously. For this purpose, many existing architectures may need to be reinvented. First of all, hosting more qubits in a limited space requires reducing the qubit footprint. Recent developments show that the shunt capacitor can be miniaturized by 100 folds or more using two-dimensional materials while maintaining coherence [328–330]. Even if the qubits can be densely packed up on a chip (size: $L \times L$), it is nevertheless extremely difficult to route all the control wires from the perimeter (length $\propto L$) to individual qubits (density $\propto L^2$) due to the unmatched scaling law; let alone to avoid crosstalk between wires. In recent years, there have been substantial efforts to exploit the third dimension to relieve this pain with various technologies borrowed from the semiconductor chip packaging, such as flip-chip bonding and through-silicon vias [331–338]. Aside from expanding the space for wiring, a different approach is to reuse the wire for multiple targets. Signal multiplexing and control line sharing schemes can alleviate the problem for future large-scale devices [304, 339, 340]. They also help reduce the cable density inside and out of the dilution refrigerator.

In the future, we may end up with insurmountable engineering challenges, including available wafer size, device yield, and crosstalk, all constraining the scalability of monolithic quantum processor designs. This suggests the desirability of developing alternative modular approaches, where smaller-scale quantum modules are individually constructed and calibrated, then assembled into a larger architecture using high-quality quantum coherent interconnects [341–345]. Several recent experiments have demonstrated deterministic quantum state transfers (QSTs) between two superconducting quantum modules, with interconnects provided by commercial niobium-titanium (NbTi) superconducting coaxial cables [346–348], copper coaxial cables [349], and aluminum waveguides [350], showing fidelities up to $\sim 80\%$, primarily limited by lossy components including connectors, circulators, and printed circuit board traces. More recent efforts using wirebond [172] or clamped [351] connections between the quantum modules and the superconducting cables have eliminated the need for normal-metal components, improving cable quality factors to $\sim 5 \times 10^4$ and inter-module QST fidelities to $\sim 90\%$ (Fig. 3h). Flip-chip modular approaches have also been pursued [352], where the qubits on separate chips are closely spaced and directly coupled, achieving high fidelities while retaining many of the benefits of a modular architecture.

In addition to the architectural design of quantum processors, supporting technologies are also crucial for scaled implementations. As a result of non-ideal fabrication conditions, the critical current of a Josephson junction usually varies by a few percent, equivalent to a few hundred megahertz in terms of qubit frequency; such unpredictable variation severely affects the quality of qubit operations. Techniques for improving junction uniformity during and after fabrication may open up new possibilities in hardware and software design [353–356]. Moreover, the capacity of a dilution refrigerator will ultimately be limited by its cooling power. Promising solutions include a careful wiring plan [357] and energy-efficient cryogenic electronics [358–360]. Simultaneous high-fidelity operations require low crosstalk. Crosstalk mitigation is a must-do. Besides optimization through design and packaging [216, 272, 361–365], various control techniques have been developed to reduce different kinds of crosstalk such as microwave signal crosstalk [279, 366], spectator effect [367–369], and residual $ZZ$ interactions [370, 371]. In the future, by integrating together various technologies in a large system, more powerful quantum processors based on superconducting qubits can be anticipated.

## IV. TRAPPED-ION QUBITS

*Introduction.*— Trapped-ion systems are the leading physical platform in pursuit of fault-tolerant quantum computing. Laser-cooled atomic ions confined in an ultra-high vacuum environment are well isolated from noises, being able to encode high-quality qubits into a stable pair of electronic energy levels in each ion, as shown in Fig. 2(b). Ion qubits hold the longest coherence time beyond any other systems [372–374], and can be initialized and measured with extremely high fidelity [372, 375, 376]. Quantum logical operations are typically performed through tailored laser or microwave-ion interactions, and gate fidelities achieved experimentally on a few ion qubits have gone well beyond the fault-tolerant threshold [372, 377–379]. Ion-based quantum processors can already primely manipulate dozens of qubits [380, 381], and quantum algorithms such as Shor's algorithm and Grover's search algorithm have been exemplified in small-scale systems [382, 383]. Meanwhile, quantum simulators with up to 53 qubits have been demonstrated to study various novel features of complex many-body systems represented by quantum spin models [384–392]. The above progress illustrates the great potential of utilizing large-scale trapped-ion systems for ultimate quantum computing.

*Ion qubit.*— Despite hundreds of atomic species that exist in nature, hydrogen-like ions are preferred for trapped-ion quantum computing due to their relatively simple atomic structures. Alkaline-earth ions like $Be^+$ [377], $Mg^+$ [393], $Ca^+$ [378, 379, 381, 394], $Sr^+$ [395], $Ba^+$ [396, 397], and rare-earth ions like $Yb^+$ [380, 398–400] are the most frequently utilized in current research. A qubit can be encoded into a pair of energy levels of a single ion, and a representative encoding scheme employs a combination of levels belonging to the ground manifold $^2S_{1/2}$ and the long-live metastable manifold $^2D_{5/2}$. This scheme is the prior choice for even-mass isotopes, such as $^{40}Ca^+$ [381] and $^{88}Sr^+$ [395], and these qubits have energy gaps on the order of optical frequencies; therefore, they are denoted as optical qubits. Although the typical lifetimes of metastable levels can reach a few seconds or so, they would ultimately limit the coherence time of optical qubits.

For odd-mass isotopes, the encoding scheme is to utilize the hyperfine splittings of the ground manifold $^2S_{1/2}$ induced by non-zero nuclear spins. The lifetime of ground-state hyperfine levels can approach the age of the universe, resulting in an extremely long relaxation time ($T_1$) compared to that of the optical qubits. Meanwhile, a certain pair of hyperfine levels can form the so-called "clock state" under a suitable external magnetic field, and its energy gap is insensitive to the static magnetic field to the first order, thus having a relatively long coherence time ($T_2$). It is experimentally observed that the $T_2$ time of a single $^{43}Ca^+$ ion can reach 50 s [372]. This record is then extended to 10 minutes in a $^{171}Yb^+$ ion qubit by using DD pulses and sympathetic cooling assisted by a $^{138}Ba^+$ ion [373]. Most recently, the one-hour coherence time has even been approached by further reducing the potential noise from the external magnetic field and the leakage from microwave sources [374]. Such a long coherence time allows systems to execute millions of gate operations before losing quantum features.

The cycle transition of $^2S_{1/2} \leftrightarrow {}^2P_{1/2}$ facilitates the realization of extremely low state-preparation-and-measurement (SPAM) errors on ion qubits. Qubit state initialization is achieved by optical pumping techniques. By choosing proper polarizations and frequencies of pumping lasers, a certain energy level of the qubit can be a dark state, and then the ion state would be pumped to this level with high probability and high speed. A typical initialization process can take a few microseconds, and infidelity can be suppressed close to $10^{-4}$ [372].

State measurement is implemented by resonating one of the qubit levels to a short-live manifold and collecting the corresponding fluorescent photons simultaneously. Projected qubit states in a single shot can be distinguished by determining whether the number of collected photons reaches a certain threshold. Depending on the photon collection rate, the measurement duration could vary from several microseconds to milliseconds, while the error can reach below $10^{-3}$ [372, 375, 376]. This error can be suppressed to around $10^{-4}$ for ions with long-live levels for state shelving [401]. Several other methods, such as adaptive analysis or time-stamping of arriving photons [375, 401, 402], are employed to further reduce measurement infidelity or increase detection speed, and meanwhile, machine learning methods can be utilized for multiple-qubit detection to reduce crosstalk errors [403, 404].

*Quantum gates.*— Quantum gates on ion qubits are typically performed by interacting ions with external laser or microwave fields, depending on the qubit encoding schemes. For example, single-qubit rotations on optical qubits can be applied through optical quadrupole transitions induced by a narrow linewidth laser, and that on hyperfine qubits can be realized by using microwaves or stimulated two-photon Raman transitions. Error rates below $10^{-4}$ have been reached for either quadrupole transitions on optical qubits or Raman transitions on hyperfine qubits [377, 378]. Error rate of $10^{-6}$ was even achieved on microwave-driven hyperfine qubits [372].

Although high-fidelity single-qubit rotations are readily accessible experimentally, the qualities of current quantum processors are mainly limited by the performance of entangling operations. The first proposal of the two-qubit gate on ion qubits, the Cirac-Zoller gate [6], is challenging to scale up due to the stringent requirements for ground-state cooling and sensitivity to thermal excitation on motional modes. However, this proposal inspired the idea of using collective motional modes of ion-chain to engineer effective qubit-qubit cou-plings. The entangling schemes used today can be categorized into Mølmer-Sørensen gates [411, 412] and light-shift gates [413, 414], which both rely on the notion of state-dependent forces. These schemes show excellent performance in experimental demonstrations. The error rate of the Mølmer-Sørensen gate below $8 \times 10^{-4}$ is achieved with two $^9Be^+$ ions [377], and that of the light-shift gate below $9 \times 10^{-4}$ is realized on two $^{43}Ca^+$ ions [378]. Recently, light-shift gates on optical qubits are theoretically investigated and then experimentally demonstrated in a two $^{40}Ca^+$ ions system [379, 415]. The gate infidelity, as low as $6 \times 10^{-4}$, is approached, representing the best entangling gate achieved ever.

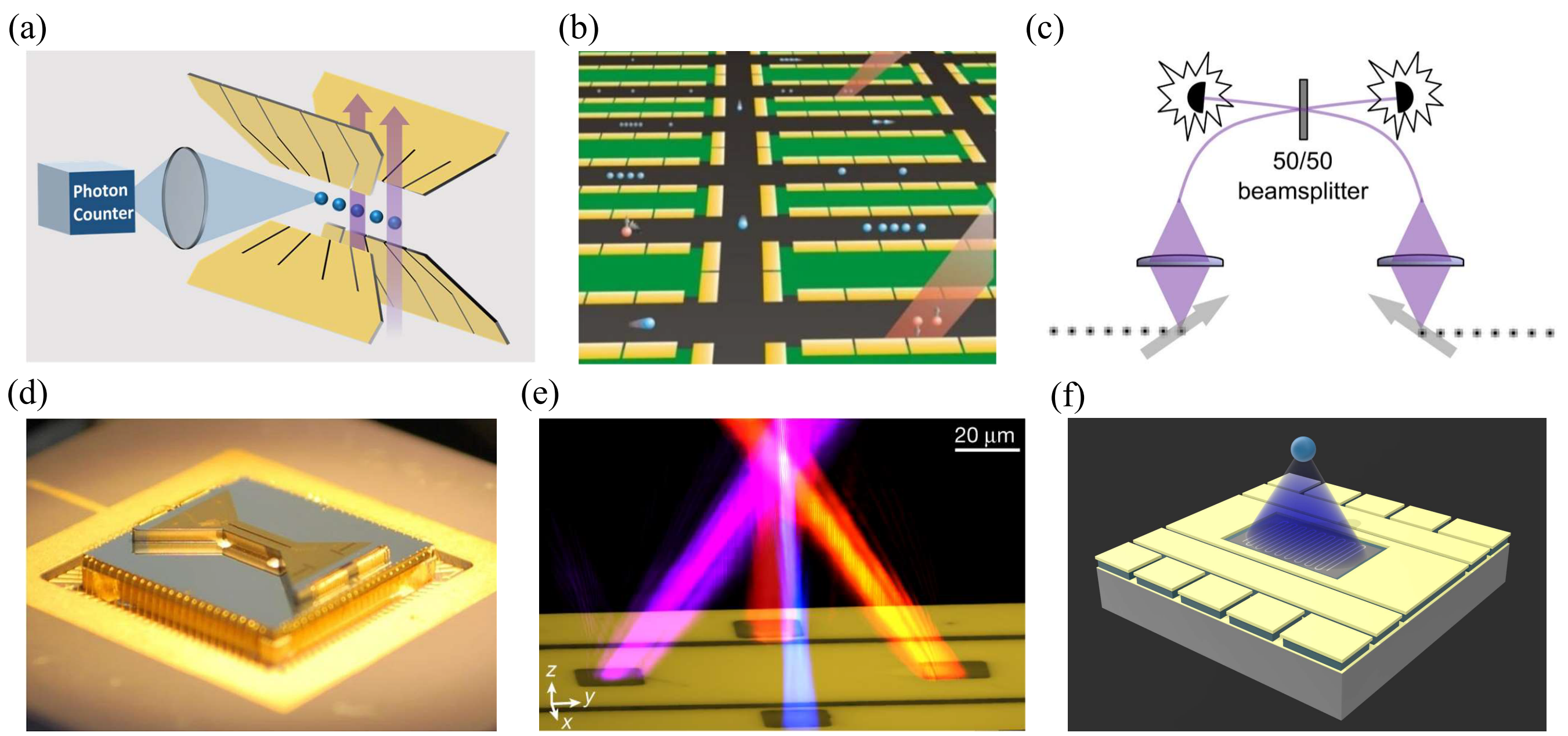

FIG. 4. (a) Three-dimensional Paul trap and captured long ion chain. Compared to the conventional four-rod trap, electrodes shown here are transformed into blade-shape to enhance optical accessibility. A one-dimensional chain of ions is trapped along the null line of the radiofrequency field, and a tightly focused laser beam array individually controls ion qubits. (b) QCCD architecture, adapted from Ref. [405]. Large numbers of ions are distributed in large chip-type traps with multiple trapping zones. Ions can be manipulated independently in different functional zones to realize logical operations, storage, or readout. Quantum information can be interchanged by transporting ions between zones. (c) Remote-ion entanglement, adapted from Ref. [405]. Ions in different traps can be heraldedly entangled by generating ion-photon entanglements and then applying Bell measurement on photons. It paves the way for large-scale distributed systems. (d) Surface trap fabricated by Sandia National lab, adapted from Ref. [405]. (e) Integrated photonic system to deliver laser beams to ions position. Here we show the surface trap with multi-wavelength integration done by the MIT group, which is adapted with permission from Ref. [406], Springer Nature Limited. (f) On-chip detection of ion qubit. Several groups have been successfully demonstrated integrated single photon detector on the fabricated surface traps [407–409].

TABLE II. Selected state-of-the-art performance on ion qubits [a].

| Qubit type | | Hyperfine qubit | Optical qubit |
|---|---|---|---|
| $T_2$ | | 50 s [372]<br>5500 s [374][b] | 0.2 s [410] |
| SPAM error | | $6.9 \times 10^{-4}$ [376] | $8.7 \times 10^{-5}$ [401] |
| 1Q gate | Duration | 1-10 $\mu$s typical | 1-10 $\mu$s typical |
| | Fidelity | 0.99996 [377] | 0.99995 [410] |
| 2Q gate | Duration | 10-100 $\mu$s typical | 10-100 $\mu$s typical |
| | Fidelity[c] | 0.9991 [377]<br>0.999 [378] | 0.9994 [379] |
| Maximally entangled qubits | | 24 [381] | |
| Environment | | Ultra-high vacuum $< 10^{-11}$ Torr | |

[a] Here we only include data from peer-reviewed publications.
[b] With dynamical decoupling.
[c] Two-qubit gate fidelities are estimated from the fidelities of the prepared Bell states.

Laser fields are mostly utilized to drive entangling gates on ion qubits due to their large spatial gradient of electric fields to provide efficient ion-motion couplings. However, microwave-driven entangling gates are also pursued due to the extreme stability of long-wavelength microwaves [416, 417]. Ion-motion coupling induced by microwave fields can be achieved by placing magnetic field-sensitive qubits into static magnetic fields with large spatial gradients or by exploiting near-field oscillating microwaves. The former scheme suffers from short coherence times induced by fluctuating magnetic fields, which can be overcome by utilizing microwave-dressed qubits [418, 419], while the latter one requires microwave sources close to the ions; therefore, the crosstalks should be taken care of. Experiments have demonstrated gate fidelities of about 98.5% [398] and 99.7% for each scheme [420, 421] respectively. Moreover, a recent experimental work has shown that with the microwave-driven laser-free gate, an almost perfect symmetric Bell state has been generated [393]. These advances promise a scalable way to achieve ion-based quantum computing with full microwave control [422].

However, most experimentally implemented entangling operations so far are relatively slow, usually on the order of tens to hundreds of microseconds, thus limiting the core speed of ion-based quantum processors. Therefore, fast gate implementation becomes one important topic of research in recent studies. The straightforward way to speed up entangling gates is to increase the laser power to enhance the laser-ion coupling strength. Along this routine, an entangling gate with a duration of 1.6 $\mu$s is achieved while the fidelity is still maintained at 99.8% [423]. However, the gate fidelity drastically drops to around 60% when the gate duration further reduces to 480 ns, due to the breakdown of the Lamb-Dicke (L-D) approximation. It might be solved by considering high-order qubit-motion couplings [424]. Another way to achieve fast gates is employing a sequence of ultrafast laser pulses to impose ultrafast state-dependent kicks on ion qubits [425–427], which does not require the ions to remain in the L-D regime. A Bell state with 76% fidelity is prepared within 1.96 $\mu$s in a recent experimental demonstration, and the main infidelities come from the imperfect kick control and off-resonant coupling to undesired energy levels [428]. High repetition-rate pulsed lasers can be helpful to further improve the gate speed [429]. Although recent implementations of fast gates still have limited fidelities, these schemes all show well scalability.

*Scalability.*—A straightforward way to scale up ion-based quantum processors is to trap multiple ions in a linear array, as illustrated in Fig. 4 (a). By exploiting the collective motional modes of the entire ion chain, entangling gates can be applied to any two ion qubits by coupling to single or multiple motional modes. For the latter case, time-modulated state-dependent forces are required to decouple multiple motional modes from ion qubits simultaneously to guarantee high-fidelity operations [430–436]. Along this route, up to 14 $^{40}Ca^+$ ion qubits were first used to generate the Greenberger–Horne–Zeilinger (GHZ) states [437], and then the qubit number was increased to 24 in a recent report [381]. Meanwhile, a programmable trapped-ion quantum processor was implemented in 2016 [438], consisting of 5 individually controlled $^{171}Yb^+$ ion qubits, and later this system was extended to 11 qubits in 2019 [380]. So far, multiple research groups around the world have realized their quantum processors with long ion chains [395, 399].

One distinct advantage of using an ion chain is the so-called full connectivity [439], which, as already mentioned, allows ion qubits confined in the same potential to be directly entangled even if they are not spatially adjacent. This feature makes the decomposition of quantum circuits more efficient and makes it possible to realize multi-qubit entangling gates. Several theoretical works have pointed out that multi-qubit gates might bring polynomial or even exponential speedups to running quantum tasks [440–443]. Therefore, researchers have been eager to explore scalable ways to achieve multi-qubit gates in recent years [399, 444, 445]. However, this linear-chain architecture also has several drawbacks, making it hard to reach a large scale. For example, the laser power required to entangle the ion qubits in a chain would increase as the size of the chain enlarges. The long chain's cooling also becomes imperfect, while gate operations become more sensitive to external noises.

To further scale up to larger numbers of qubits, we can trap multiple ion chains in several independent potentials and construct some link channels for interconnections. One representative architecture is the quantum charged-couple device (QCCD) proposed in 2002 [445] (see in Fig. 4 (b)). Links between chains are achieved by modifying local electric potentials to redistribute ions between trap regions physically. To achieve this goal, shuttling operations [446] like linear transport, splitting or merging of ion-chain, and position swap should be included together with quantum gates in local chains as the basic operations of quantum processors. These operations must be performed fast enough so that they do not become processing speed bottlenecks. Several fast shuttling methods have been fully investigated and demonstrated to simultaneously satisfy these two requirements [447, 448], promising to construct a reliable highway of ion qubits to enable a large-scale QCCD architecture.

However, the complexity of the ion trap is significantly increased compared to that used for a single linear chain. An ion trap with numerous independent control electrodes is required to realize multiple trap regions and precise control of the ion shuttle. Microfabricated chip traps are a satisfactory solution to these complexities [449, 450]. Recently, a series of 1D chip traps and 2D traps with X-type [451], Y-type [452], or T-type [453] junctions have been presented. Utilizing these well-controlled traps, 4 qubits GHZ state has been prepared in a shuttling-based way [394], and quantum gate tele-portation has also been demonstrated [454]. Moreover, high-quality quantum processors based on QCCD architecture have shown excellent performance according to quantum volume measurement [455]. However, towards large-scale surface traps, there are still several challenging issues that should be addressed. One critical problem is the anomalous heating on motional modes of ion chains [456–458]. It would destabilize ion chains and become one of the main error sources for quantum gates and ion shuttling. Although many efforts have been made to reveal the origins of this heating effect, the problem is still not well solved. Other issues such as radio frequency (RF) potential barrier [459] in the junction transport and relatively low trap depth would also plague the development of scalable QCCD-based quantum processors. Nevertheless, QCCD architecture is still an outstanding approach for large-scale trapped-ion quantum computing.

Another natural choice to link ion qubits in different regions is utilizing photons [460]. By exciting ion qubits to a short-live ancilla atomic level, the polarization of the spontaneously emitted photon would entangle with a decayed ion state. By applying Bell measurement to the photon pair from different ions, a heralded entanglement can be generated between ion qubits depending on the measurement outcomes, as shown in Fig. 4 (c). This remote-ion entangling method enables the feasibility of entangling ion qubits in different vacuum systems even if they are far away, leading to the paradigm of distributed quantum computing. The generating rate of the remote entanglement should be fundamentally determined by the scattering rate of the ancilla level. However, in practice, this rate is mainly limited by the collecting rate of the emission photons. In recent experimental demonstrations, a generation rate of 4.5 Hz [461] is first realized and then improved to 182 Hz [462], while the best fidelity of the heralded qubit entangling state is 94%. This value is much slower than the gate speed of directly entangling qubits in the same trap. Several methods have been proposed to further increase the generation rate, such as enlarging the numerical aperture of the photon collecting system and increasing the quantum efficiency of the single photon counter. Significant improvement might be achieved if we place ion qubits into a high finesse micro-cavity, enhancing the spontaneous emission through the Purcell effect [463]. Moreover, the conversion of a single photon from visible regime to telecom-wavelength has been demonstrated recently [464], although with quite low converting efficiency, making it possible to build distributed quantum systems with ultra-low optical loss. Practical remote-ion entanglement would facilitate the construction of large-scale quantum computing platforms and also quantum networks.

Here we briefly talk about hybrid-ion systems. When we generate remote entanglement of two traps or apply mid-circuits measurement to certain ion qubits belonging to an ion register, we want to only excite targeted ion qubits without disturbing others resonantly. One solution is that the measured ion and others belong to different species, so the resonance frequency is quite different, significantly suppressing crosstalk. Consequently, entangling gates on ion qubits of mixed species are required in hybrid-ion systems. High-fidelity entangling operations on mixed-species ion qubits are well displayed on such as $^{9}Be^{+}$–$^{25}Mg^{+}$ [454], $^{40}Ca^{+}$–$^{43}Ca^{+}$ [465], $^{43}Ca^{+}$–$^{88}Sr^{+}$ [466], $^{171}Yb^{+}$–$^{138}Ba^{+}$ [467, 468] and even a long chain of $^{9}Be^{+}$–$^{9}Be^{+}$–$^{40}Ca^{+}$ [469]. Meanwhile, mixed-species systems make it feasible to apply sympathetic cooling, allowing to cool down ion register without destroying stored quantum information [470–472]. It is valuable for suppressing motional excitation during ion shuttling in the QCCD architecture. Meanwhile, proposals for hybrid encoding, by utilizing multiple energy levels of a single ion to encode different qubit types, have been made recently to construct hybrid systems with single ion species [473]. The interconversion of different qubit encoding on a single ion has been experimentally demonstrated [474]. It might open a new way for scalable trapped-ion quantum computing.

*System integration.*— System integration is inevitable in building scalable large-scale trapped ion quantum computers. One typical example is the microfabricated surface traps mentioned above. In the past decade, researchers have done several important works to promote the integration of trapped ion systems further, and on-chip integrated optics is one of them. Laser beams can be delivered and tightly focused at the location of the ions by embedding optical waveguides beneath the surface traps and fabricating properly designed gratings coupler at the end of each waveguide. Optical integration from single to multiple wavelengths has been well-demonstrated [406, 475]. Single qubit rotations and two-qubit entangling gates have been implemented using laser beams delivered through waveguides [475, 476], showing extreme robustness against vibrational noises. Furthermore, the integration of single photon detectors on chip traps has been successfully demonstrated recently [407–409], showing a scalable way for high-fidelity readout of multiple ion qubits on large-scale quantum processors. Conventional analogue voltage sources are also integrated on-chip [477], enabling an expandable approach to control numerous ion trap electrodes and laying the technical foundation for circuit integration in large-scale QCCD architectures.

*Outlook.*— This section briefly reviews the significant advancements in trapped ion quantum computing over the past decades, from excellent control of several ion qubits to demonstrations of scalable architectures. With fully controllable trapped-ion processors, several great advances have been made recently in QEC [410, 469, 478–481]. However, we still strive to explore specific scalable ways to achieve a truly fault-tolerant logical qubit encoding and ultimately build an applicable ion-based quantum processor. One of the fundamental requirements is that the size of the system increases without compromising the quality of the control [482–484]. Therefore, developing system integration-related technologies would be critical for scaling-up ion quantum processors in the following decades. Meanwhile, by merging the techniques developed for trapped-ion quantum computing, we might also gain better performance in ion-based precision measurement. With continuous development, trapped-ion systems would remain an important platform and tool for future quantum information applications.

## V. SEMICONDUCTOR SPIN QUBITS

*Introduction.*— Spin qubits in semiconductors have made tremendous progress over the past few decades. Although most toolboxes have been built based on GaAs quantum dots [491], this field had a revival after the host material steered to silicon. Further momentum is gained after some recent vital breakthroughs, such as fault-tolerant quantum gates [492–494], rf-reflectometry spin readout [495–498], spin-photon strong coupling [499–503], hot qubits [504–506], and cryo-CMOS controlling chip [507, 508]. Combined with its inherent scalability from the semiconductor industry and its small footprint, spin qubits are now well poised for the following milestones—quantum advantages over classical supercomputers, prototype machines of fault-tolerant quantum computing and QEC, and hybridization between classical and quantum electronics.

*Qubit construction.*— Semiconductor qubits are defined on the charge and spin degrees of freedom of the carriers trapped in quantum dots or dopants. There are several types of qubits, such as spin qubits [10, 11, 509, 510], charge qubits [511, 512], exchange-only qubits [44, 513, 514], hybrid qubits [515], and singlet-triplet qubits [516, 517]. A spin qubit is usually defined by the spin states of a single electron or hole trapped in a semiconductor quantum dot or a dopant in silicon (see Fig. 5) [10, 11], as $|0\rangle = |\downarrow\rangle$ and $|1\rangle = |\uparrow\rangle$, where $|\downarrow\rangle$ and $|\uparrow\rangle$ denote spin down and spin up, respectively. Chosen states of an interacting multi-spin system can also be defined as a qubit, such as the singlet-triplet qubit (i.e., the singlet state $|S\rangle = (|\uparrow\downarrow\rangle - |\downarrow\uparrow\rangle)/\sqrt{2}$ and the unpolarized triplet state $|T_0\rangle = (|\uparrow\downarrow\rangle + |\downarrow\uparrow\rangle)/\sqrt{2}$ of two exchange-coupled spins). Here we will focus on the spin qubits in this review, as they are becoming the central topic in the field in recent years. We will also briefly introduce other types of qubits. Interested readers can find more information in the relevant references [491, 509, 510, 518–520].

*Decoherence.*—The development of semiconductor host materials and related fabrication technologies underpins progress in this field. GaAs/AlGaAs heterostructure quantum well has been the key substrate for gate-defined quantum dots [529], where this field has accumulated many of its foundational building blocks, such as single charge sensing, single-shot spin readout, qubit operations, and interactions, to mention only a few. Nevertheless, the nucleus of the host GaAs forms a fluctuating magnetic field, namely the Overhauser field, which limits the spin dephasing time $T_2^*$ in GaAs to around the range of tens of nanoseconds [491, 510, 529]. Although GaAs quantum dots still stand out nicely as a demonstration platform for quantum simulation [530, 531] and quantum physics research [532, 533], the limited spin dephasing time hinders the development of high-fidelity quantum gates for quantum computing.

A host material consisting of zero nuclear spin isotopes is necessary to embrace a long dephasing time. In his seminal paper 20 years ago, Bruce Kane emphasized that group IV materials are ideal options as they feature stable zero-nuclear spin isotopes with high natural abundance [10]. The nuclear-spin free isotope $^{28}$Si, for example, could be accessed via purifying the natural silicon. Another advantage of the silicon substrate is the long spin $T_1$ times, which comes from the different spin relaxation behavior compared with the group III-V materials [510]. Since $T_1$ sets an upper bound on $T_2$ for a spin system via the relation $2T_1 \geq T_2$, a long $T_1$ is the prerequisite for having a long $T_2$. Therefore, silicon spin qubits have become the working horse in this field, and the mainly explored platforms include silicon metal-oxide-semiconductor (MOS) [489, 523, 525], silicon-on-insulator (SOI) [488, 534], the dopant in silicon [490, 525, 535], silicon-germanium heterostructures based Si/SiGe [492, 524, 536, 537], and Ge/SiGe [538–541]. After extensive studies, competitive $T_2$ and $T_2^*$ compared to the other major quantum computing platforms are demonstrated in silicon spin qubits. Up to now, $T_1$ values ranging from 160 milliseconds [522] to 30 seconds [542] and $T_2^{\mathrm{Hahn}}$ values ranging from 99 microseconds [524] to ∼1 second [489] are observed in silicon dopant devices, Si/SiGe gate-defined systems, and Si-MOS systems (for detailed comparison, please refer to Table III). Other silicon spin systems also exhibit typical long coherence times, such as hole spins in SOI [488] and Ge/SiGe gate-defined quantum dots [539]. Long coherence sustains even at 1.1 K∼4.2 K in MOS [504, 505] and SOI devices [506].

*Gates.*—Over the past years, prominent progress has been made for quantum gate fidelities in semiconductor qubits. A single-qubit operation could be performed utilizing a few different mechanisms for different qubits. For singlet-triplet qubits [517, 543], exchange-only qubits [44, 513, 514], and hybrid qubits [515], exchange coupling plays the key role. In comparison, electron spin resonance (ESR) [525, 544, 545] and electrical dipole spin resonance (EDSR)[524, 538, 546, 547] are the main driving mechanisms for a spin qubit. They control the spin rotation by either oscillating magnetic or electrical fields. After years of persistent quest for fault-tolerant operations, single-qubit gate fidelities well beyond 99% have all been demonstrated in the donor [489], Si-MOS [485], and Si/SiGe [486, 514, 524, 548] (see Table III).

Compared to single-qubit gate operations, two-qubit gates are more daunting. Exchange coupling, capacitive coupling, or hyperfine coupling can be utilized to realize a two-qubit gate. In the singlet-triplet qubit, capacitively coupled two-qubit gates have been shown [549, 550].

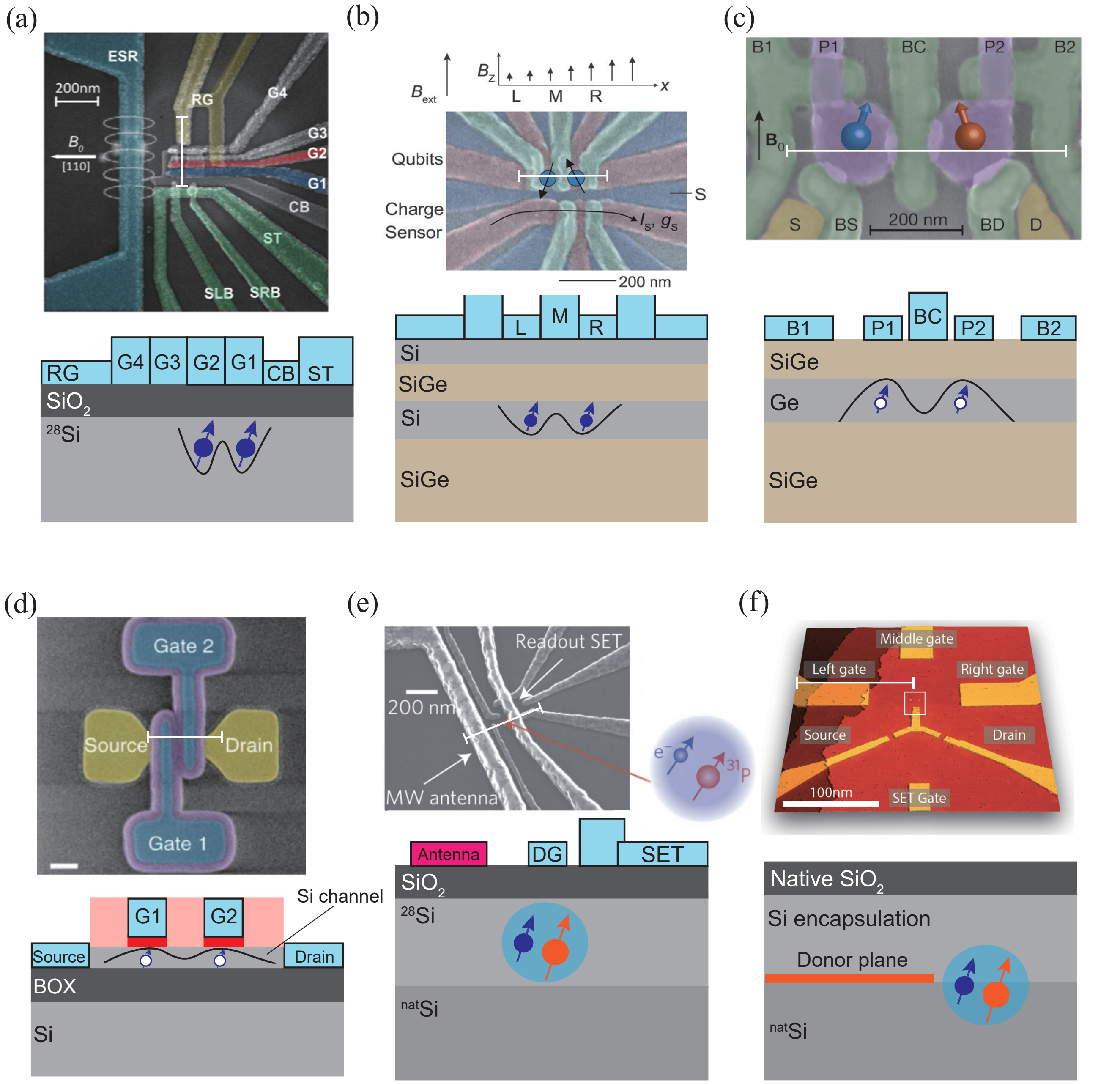

FIG. 5. Representative semiconductor qubit systems. All the devices are presented with two panels, where a top panel shows the top view of the device, and a bottom panel shows the lateral structure corresponding to a white line cut of the active region in the top panel. Heterostructure quantum dots include (a) Si-MOS [485], (b) Si/SiGe [486], and (c) Ge/SiGe [487] systems, where Si-MOS and Si/SiGe are mainly used for electron qubits, and Ge/SiGe is a hole qubit platform. System **(a-c)** belong to the gate-defined quantum dot category. In (a) Si-MOS, quantum dots are formed close to the silicon-oxide interface, with fabricated top gates providing lateral Coulomb confining potentials. On this device, an electron spin driving ESR antenna and a spin readout single electron transistor are integrated as well. (b) Si/SiGe quantum dots are formed in the middle silicon quantum well layer and sandwiched between SiGe layers on both sides. (c) Ge/SiGe is a hole spin platform, using a Ge well to form a two-dimensional hole gas and combined with top gates to form hole quantum dots. (d) CMOS nanowire field-effect transistor [488], where quantum dots are formed in the silicon nanowire sitting on a buried silicon oxide layer (BOX), and the surrounding gates are fabricated using industrial microelectronics technology. It is a hole-spin system, and the dot potential well is formed at the valance band top. (e) and (f) are the donor qubits in silicon. They are electron spin systems, where the host donor nuclear spin is also a key resource to encode qubits. (e) represents a donor in MOS [489] system, where a phosphorus atom is implanted in a fabricated MOS device which has a $^{28}$Si layer to prolong electron spin coherent time. (f) Donor device with STM lithography technique [490], where the donors can be placed with atomic precision and in-plane gates are formed by dense conducting phosphorus atoms in a lithographical manner. Top panel of (a) is adapted with permission from Ref. [485], Springer Nature Limited. Top panel of (b) is adapted with permission from Ref. [486], American Association for the Advancement of Science. Top panel of (c) is adapted with permission from Ref. [487], Springer Nature Limited. Top panel of (e) is adapted with permission from Ref. [489], Springer Nature Limited.

TABLE III. Comparison of reported values of different qubits in silicon.

| Qubit type | Si-MOS | Si-SiGe | P donor n | P donor e |
|---|---|---|---|---|
| $T_1$ | 2.6 s [521] | 160 ms [522] | 39 min [489] | 30 s [489] |
| $T_2^*$ | 120 $\mu$s [523] | 20 $\mu$s [524] | 600 ms [489] | 268 $\mu$s [489] |
| $T_2^{\mathrm{Hahn}}$ | 1.2 ms [523] | 100 $\mu$s [524] | 1.75 s [489] | 0.95 ms[489] |
| $T_{\mathrm{single}}$ | 2.4 $\mu$s [523] | 20 ns [524] | 24 $\mu$s [493] | 150 ns [525] |
| $T_{\mathrm{two}}$ | 1.4 $\mu$s [485] | 103 ns [492] | 1.89 $\mu$s [493] | 0.8 ns [490] |
| $F_{1\mathrm{RB}}$ (%) | 99.957(4) [526] | 99.861(5) [524] | 99.99 [527] | 99.95 [527] |
| $F_{2\mathrm{RB}}$ (%) | 98.0(3) [485] | 99.51(2) [492] | 99.37(11)[a][493] | 86.7(2)[b][490] |
| $Q_1$[c] | 50 | 1000 | 25000 | 1800 |
| $Q_2$[c] | 86 | 194 | 302[d] | $3.4 \times 10^5$ |
| $N_{\mathrm{Q}}$[e] | 2 [485] | 6 [528] | 2 [493] | 2 [490] |
| $N_{\mathrm{E}}$[f] | 2 [485] | 3 [528] | 2 [493] | 2 [490] |
| Env | $B \sim 1.4$ T<br>$T < 1.5$ K | $B \sim 0.5$ T<br>$T < 1.5$ K | $B \sim 1$ T<br>$T < 1.5$ K | $B \sim 1$ T<br>$T < 1.5$ K |
| Flying qubit | N/A | N/A | N/A | N/A |
| Footprint size | $\sim 100$ nm | $\sim 100$ nm | $\sim 3$ nm | $\sim 100$ nm |

[a] CZ gate.
[b] $\sqrt{\mathtt{SWAP}}$ gate.
[c] $Q_1 \equiv T_2^*/T_{\mathrm{single}}$ and $Q_2 \equiv T_2^*/T_{\mathrm{two}}$, where $T_{\mathrm{single}}$ and $T_{\mathrm{two}}$ are the time for the single-qubit and the two-qubit operations.
[d] $T_2^*$=570 $\mu$s [489] is used here for a P nuclear spin with a bounded electron.
[e] $N_Q$ is the demonstrated number of qubits with individual control.
[f] $N_E$ is the number of entangled qubits.

Capacitively coupled two-qubit gates have been realized in charge qubits as well [512]. In spin qubits, the well-established two-qubit gate protocols, such as SWAP [490, 516], C-Phase [523], and C-ROT [486], all need the existence of exchange coupling. Since either the capacitive or exchange coupling hinges on the charge degrees of freedom, charge noise couples into the system and renders a challenge for high-fidelity gate operations. Different methods have been pursued to realize high-fidelity two-qubit gates. A fixed exchange coupling was used for Si/SiGe quantum dots [492], where the unwanted rotation of the off-resonant states was removed by carefully matching the Rabi frequency $f_R$ and the exchange J relation, and a 99.8% fidelity two-qubit gate was realized. The other two teams both utilized the tunability of the exchange coupling to perform CZ gates in the Si/SiGe platform. After detailed calibration and pulse optimization, two-qubit gate fidelities of $F_{\mathrm{CZ}} = 99.65\%$ [494] and $F_{\mathrm{CZ}} = 99.81\%$ [551] were demonstrated. In the silicon donor system, a two-qubit CZ gate with $F_{\mathrm{CZ}} = 99.37\%$ is shown on two donor nuclei with a shared electron [493] using a geometric gate and hyperfine coupling.

Despite of the milestone breakthroughs in the fault-tolerant qubit gates, the semiconductor qubits still have much room to improve the qubit operation fidelities. Especially for the charge noise issue [524, 526, 552, 553], qubit host material engineering is necessary to have a purer environment, such as fewer nuclear spins, charge traps, and defects. Moreover, more sophisticated control methods or encoding could be combined, such as dressed qubits [554], global control [555], and profile optimized pulses [556]. Besides, the sweet spot in the qubit energy [552, 557] and composite pulse sequences [558] could also help against the charge noise. We stay optimistic about further improvements in gate fidelities from optimized device fabrication and control-level engineering.

*Readout and initialization.*—Reducing SPAM errors is as important as improving gate fidelities for pushing the fault-tolerant quantum computing. Single-shot spin readout is vital, as some state initialization protocols can be performed by just performing a readout. The key is to find a suitable state-to-charge conversion process, where Pauli spin blockade [534, 559], hence the related latching mechanism [560], and Elzerman type spin-dependent tunneling process [561] are explored. Hyperfine coupling is a key anchor for the nuclear spin readout in the dopant system where the state information can be converted to a spin ESR signal [562]. Next, the corresponding charge signal shall be correlated to the capacitive difference, picked up by a single-electron transistor (SET) or quantum point contact (QPC), and amplified further [563]. In singlet-triplet qubits [559], a readout fidelity of 98.4% is reported. For single spins, the readout fidelity is pushed to $F_{\mathrm{M}} = 99.8\%$, beyond the fault-tolerant level [542].

Single-lead rf-reflectometry spin readout [564] is a pivotal technology to reduce the gate density for spin readout and is compatible with surface code scalable architectures for the fan-out issue. It was demonstrated nearly simultaneously by four groups [495–498]. Using this technique, readout fidelities above 98% were shown [497, 498], and a readout time of 6 $\mu$s was proved [498], comparable to the gate operation speed. To remedy the broadening of the Fermi surface and related obstacles for the Elzerman protocol at high temperatures (1 K–4 K), the Pauli spin blockade, latching mechanism [539, 560], and double SET readout [565] are valuable approaches. Also, to improve the signal-to-noise ratio (SNR), quantum noise-limited JPA [566] and other amplification methods were combined. Multiple spin readout has been shown with a single electron box [567] and done by frequency multiplexing [568]. Several teams have already demonstrated enhanced readout fidelities for S-T qubits and spin qubits in a non-demolition method [569, 570], which shall find their importance in fault-tolerant computing and in studying spin state collapse problems for fundamental quantum mechanics. Moreover, cascade readout [571], dispersive spin readout [499, 534], triple-dot cavity dispersive readout [572], and ramped spin measurement [573] have been shown. In general, the spin readout techniques are more mature and ready for the future scaling-up stage. Further progress will focus on high-level multiplexing and integrating with the qubit design in a scalable manner. Potential readout signal crosstalk also needs further investigation and engineering design. To facilitate spin initialization, except the usual applied readout-assisted state initialization, a hot spot on the energy level due to valley mixing [521, 574] and spin-orbital coupling could also be used to enhance the $T_1$ relaxation rate.

*Scalability.*—An unique advantage for silicon qubits comes from its industry backbone, very large-scale integration (VLSI) technology. This advantage becomes more significant as integrated cryo-CMOS and hot qubits techniques emerge. An industry-level CMOS-compatible hybrid quantum computing chip with classical control unit and quantum processing unit is becoming possible. Along with this spirit, progress has been made in recent years, such as Intel's horse-ridge cryo-CMOS chip demonstrating qubit control fidelities rivaling room temperature bulky control instruments [507]. A proof-of-principle experiment shows a low-temperature classical control unit bonded to a quantum unit [508]. Also, on the quantum side, CMOS compatible "hot" qubits operated at increased temperature ($T > 1.5$ K) for the electron spin [504, 505] and hole spin systems [506] are realized nearly simultaneously. On the industry side, advanced fabrication technologies are touching down on qubit physics with the industrial foundry's massive production process [575]. CEA-Leti [576], Imec [577], and Intel all processed silicon qubits with 300 mm technology [578], and Intel has shown promising quantum dot uniformity and basic spin qubit operations.

In the quest for fault-tolerant computing, scaling up would be an inevitable technological hurdle to overcome [579]. Several scaling architectures have been designed for phosphorus donor qubits [557, 580], Si/SiGe gate-defined dots [581], and MOS quantum dots [582, 583]. Also, research teams have realized multi-qubit devices and few-qubit algorithms, such as three qubits [584] and six qubits in the Si/SiGe systems [528], where entanglement states were shown. Similarly, a four-qubit in the Ge/SiGe hole spin system also made its debut, where a four-qubit GHZ state was demonstrated [539]. Meanwhile, methods for multiple quantum dots tuning using virtual gates [585] and automated machine learning were developed [586, 587] and multiplexed quantum dots readout was realized [568, 588]. Toward QEC, a three-qubit phase error correction algorithm was carried out by two groups [25, 26]. Moreover, qubit networks could be a remedy for the dense packaging problem and could reduce the fan-out overhead [579]. Therefore, the coupling of spin qubits at a distance is necessary. The effective spin-spin coupling could be realized by using microwave cavities [502, 503], mediated big quantum dot [589], spin array state transfer [532], and surface acoustic waves [590]. For the cavity approach, strong coupling [591] and photon-mediated spin-spin interactions are demonstrated [502, 503]. The next step would be to realize cavity-mediated two-qubit gates for spins at a distance and hence spin networks in a distributed manner.

*Conclusion.*—High-quality materials and advanced fabrication technologies are the cornerstones of semiconductor qubits. To realize a higher gate fidelity, interfaces with low charge noise are critical, which could be achieved by importing industrial techniques of integrated circuits and encouraging a transfer from lab-level engineering to foundry-level fabrication. The typical overheads for scalable quantum computing, such as crosstalk, gate heating, and frequency crowding, should also be carefully considered. Subsequently, semiconductor spin qubits will join the other sophisticated quantum computing platforms for next-level applications, such as fault-tolerant operations and quantum simulations on intermediate-scale multi-qubit devices. In summary, semiconductor qubits, especially the silicon spin qubits, are well-positioned for scaling up with all those above-mentioned breakthroughs and technological improvements. We are confident that by embracing its industrial advantages, silicon spin systems will speedily scale up their qubit numbers and join the next-level quantum computing endeavor together with superconducting and ion trap platforms.

## VI. NV CENTERS

*Introduction.*— NV center is a point defect in the diamond, where a vacancy and a nitrogen atom substitute for two carbon atoms along the quantization axis (assumed to be the $\hat{z}$ axis), as shown in Fig. 6 (a). The negative charge state $NV^-$ is of greatest interest, where there are five unpaired electrons originating from the nitrogen atom and three carbon atoms, respectively, together with an additional electron captured from the environment. This six-electron system is equivalent to an electron with spin projection $S = 1$, whose spin state can thus be employed as a qutrit, or a qubit if only the $|m_{\mathrm{s}} = 0\rangle$ and $|m_{\mathrm{s}} = -1\rangle$ energy levels are considered. The NV center is a promising candidate for the quantum computer by virtue of the following merits. First, a single NV center can be optically resolved and located, and the polarization and measurement can also be achieved with a laser pulse. Second, the NV center has an excellent coherence property even at room temperature. At low temperature, it can be resonantly excited to enable efficient single-shot readout. Third, the nuclear spins near NV centers serve well as abundant available memory qubits for solid-state quantum information processors.

*Qubits and coherence.—* The exceptional lifetime of NV electron spins even under ambient conditions is experimentally favorable for quantum computation. The inhomogeneous magnetic fluctuations due to the $^{13}$C spin bath are the main noise source responsible for the dephasing of NV electron spins. However, with the widely used DD technique, the dephasing time can be extended from the order of microseconds ($T_2^*$) to milliseconds ($T_2$), where the quasi-static noise is mostly suppressed.

In addition to the central electron spin, nearby nuclear spins are a rich resource for memory qubits. The relatively low gyromagnetic ratio (around three orders of magnitude less than that of electron spin) of nuclear spins is mainly responsible for their extraordinarily long coherence time, up to $T_2 = 2$ s at room temperature [592]. In addition to the $^{14}$N and $^{13}$C nuclear spins [593] (up to 27 spins nowadays [594]), researchers have been endeavoring to explore more available qubits in diamond, including P1 centers [595–597] and long-lived carbon nuclear spin pairs [598] ($T_2 = 1$ min and $T_1 > 6$ min at 4 K [599]). Further improving the coherence time of NV electron spins compared to the timescale of these memory qubits is desired. Since $T_2 \sim$ ms at room temperature is limited by the spin relaxation time $T_1$, a straightforward solution is to lower the temperature, where $T_2$ has been extended to 0.6 s at 77 K [600]. At 4 K, however, $T_2$ has reached 1.5 s with carefully-designed sequences decoupling unwanted interactions, and $T_1$ has exceeded 1 h [598].

*Initialization and readout.—* NV center can be optically initialized and readout due to the spin-dependent inter-system crossing (ISC) [601, 602] (see Fig. 6 (a), illustrated by grey dashed lines). Thus, the electron spin can be initialized to $|m_s = 0\rangle$ under continuous optical pumping. On the other hand, ISC leads to a nonradiative transition through singlet states, which enables the discrimination of the spin states according to the fluorescence difference. Furthermore, at low temperature, the resonant optical excitation allows high-fidelity single shot readout of NV electron spins [603]. The preparation and readout fidelity have achieved 99.9% [604] and 98% [605], respectively.

The approach to initializing nuclear spins is less straightforward and varies with the different coupling strengths. Specifically, for $^{14}$N and some strongly-coupled $^{13}$C nuclear spins ($\geq 1/T_2^* \sim 10^2$ kHz) [606, 607], applying a well-aligned magnetic field $\sim 500$ Gauss leads to the excited state level anti-crossing (esLAC) and the polarization of nuclear spins [608]. However, esLAC fails to enable the efficient polarization if the quantization axis of the hyperfine interaction in the excited state differs from that in the ground state, or if the nuclei-electron quantization axis differs from that of the NV itself [607]. On the other hand, the strongly-coupled nuclear spins are less abundant than the weakly-coupled ones, and hence researchers have focused more on weakly-coupled $^{13}$C nuclear spins. The initialization of the weakly-coupled $^{13}$C nuclear spins employs a swap-like gate constructed by the DD sequences (see below) [28].

In addition to the approaches discussed above, several strategies exist for different purposes. For example, dynamic nuclear spin polarization (DNP) is designed to polarize the whole $^{13}$C spin bath by imposing the Hartmann-Hahn double resonance [609], and has been improved to be more robust [610]. Projective measurement-based initialization is also preferred, especially in the case of simultaneous multiqubit or multidegree-of-freedom initialization with high fidelity [597, 611]. The single-shot readout associated with projective measurements not only provides an efficient way to polarize nuclear spins, but also enables the direct test of non-classical correlations and active feedback in QEC protocols. Single-shot readout of $^{14}$N [612], weakly-coupled nuclear spins [613], nuclear spin pairs [599], and P1 centers [597] have been realized, respectively.

*Gates.—* The control techniques have been well developed to implement quantum logic gates with high precision as well as narrow pulse widths, where quantum optimization algorithms have been exploited. Combining the composite pulses with a modified gradient ascent pulse engineering (GRAPE) algorithm has yielded the records of the fidelity of single-qubit gate and two-qubit gate being 0.99995 and 0.992 [617], respectively, which almost hits the threshold value required by QEC. Remarkably, this highly accurate two-qubit gate has a duration of 700 ns, which is three orders of magnitude smaller than the coherence time. Meanwhile, typical single-qubit gates are in the order of $\sim 10$ ns, and gigahertz Rabi oscillations are also possible with proper design [618, 619]. Moreover, optimized ultra-fast single-qubit gates beyond the rotating wave approximation have been realized with the application of chopped random basis (CRAB) quantum optimization algorithm, with fidelity for $\pi/2$ and $\pi$ pulses being $0.95 \pm 0.01$ and $0.99 \pm 0.016$, respectively [620].

Manipulating multiple qubits while maintaining coherence is a crucial task for quantum computing. In particular, in hybrid systems, where the timescale of each component may differ, it is desirable to implement all the control sequences before any of the components decohere. Instead of counting on the isotopically purified samples [621], an active way to extend the coherence time is the well-known DD techniques, during which the quasi-static noise is flipped and canceled, and hence can also be construed as a frequency filter [622]. In this sense, the DD sequence applied to NV electron spins enables the detection of the resonant frequencies corresponding to surrounding interactions [623–625]. Consequently, con-ditional two-qubit gates have been designed based on DD sequences to control $^{14}$N [626], where an RF pulse acts equivalently as a transverse hyperfine coupling to drive the nuclear spin flip conditioned on the state of the electron spin. Similarly and subsequently, universal DD-based gates on weakly-coupled nuclear spins have been achieved [28, 627], through which the nuclear spins can be initialized and measured via swap-like gates. Recently, an active phase compensation scheme named DDrf has freed the dependence of interpulse delay on hyperfine parameters and enabled the optimization of the interpulse delay to protect electron coherence, eventually entangling up to seven nuclear spins in a ten-qubit register [628].

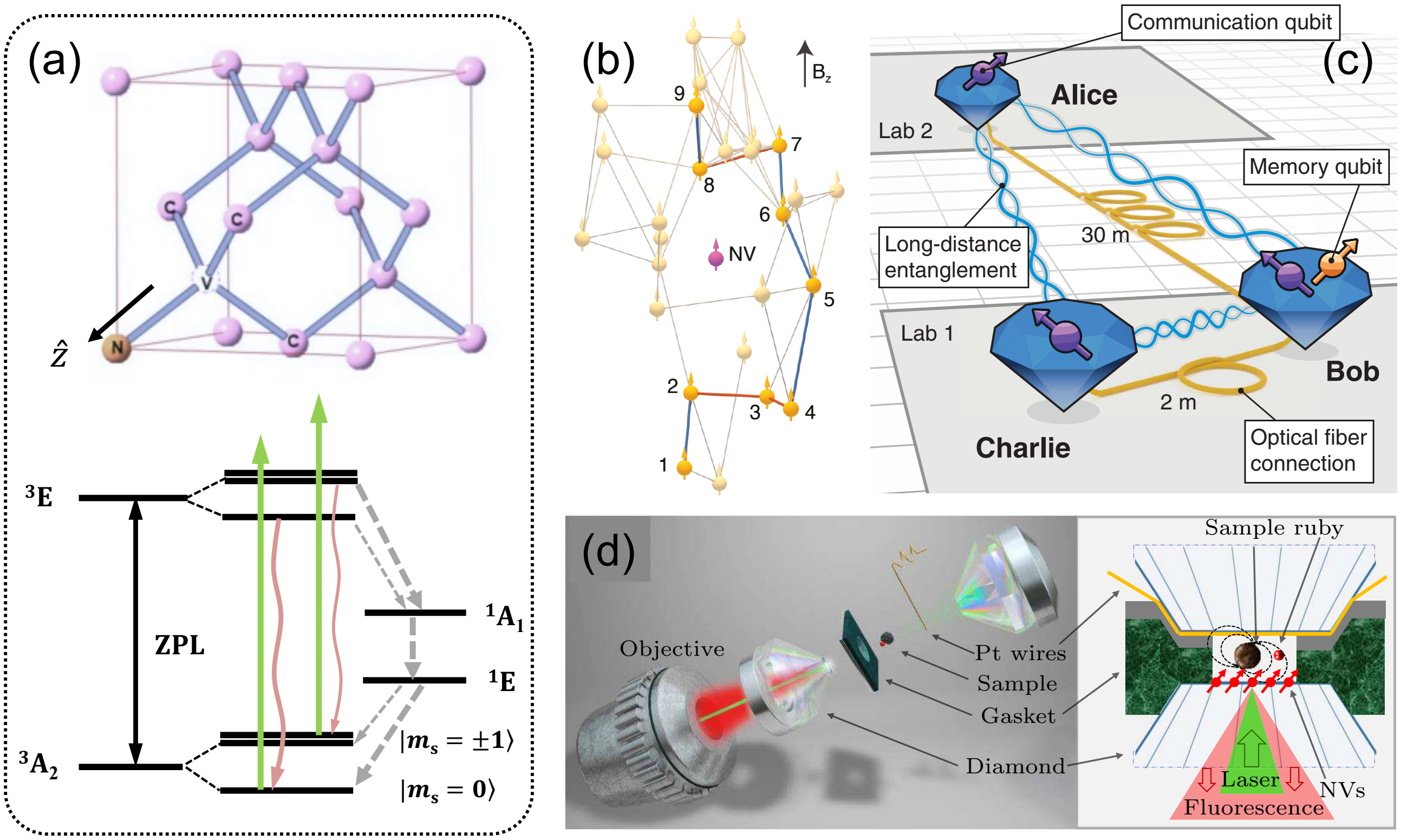

FIG. 6. (a) Physical system and level structure of the NV center. (b) NV center and the nuclear spins nearby form a multi-qubit quantum information processor [614]. (c) NV center with the nuclear spins as memory qubits form a node in the quantum network [615]. (d) NV center as a quantum sensor [616].

An alternative is to utilize decoherence-protected subspaces, where the evolutions of quantum states are purely unitary [629, 630]. Moreover, geometric gates are also intrinsically noise-resilient, where the dynamical phase vanishes. Both non-adiabatic [631–633] and adiabatic [634] universal (non-Abelian) [633] geometric gates have been demonstrated. Nevertheless, towards large-scale quantum technologies, QEC is expected to be a more essential strategy, with respect to encoding the physical qubits subjected to the deleterious noise from the environment into reliable logical qubits. Armed with state-of-the-art multi-qubit control techniques, QEC protocols [27, 28] and a related work deploying a robust coherent feedback control [635] have been realized. Most recently, fault-tolerant operations on the logical-qubit level have been achieved on a seven-qubit NV quantum processor, indicating a major step toward fault-tolerant quantum information processing [636].

*Quantum simulation and quantum algorithm.*— Sophisticated control of spins in diamond promises rich applications in diverse fields. Various exotic physical phenomena have been simulated, such as the emulations of tensor monopoles [638] and quantum heat engines [639], opening avenues for the exploration of fundamental physics. NV center quantum simulator also expands the scope of experimental investigations on quantum topological phases [640, 641]. Ref. [642] proposed a feasible and universal approach to investigate the non-Hermitian Hamiltonian in Hermitian quantum systems and observed parity-time symmetry breaking in an NV quantum simulator [637, 643]. Besides, simulations of non-Markovian dynamics of open systems [644], many-body localized discrete time crystals [614] and emergent hydrodynamics [645] are distinguished from other artificial platforms due to the real quantum nature and shed light on condensed-matter physics.

On the other hand, many quantum algorithms have also been demonstrated, including the Deutsch-Jozsa algorithm [646], adiabatic quantum factorization [647],

| Property | Parameter | Qubit | Value | Condition | Reference |
|---|---|---|---|---|---|
| **Coherence** | $T_2^\star$ | e | 36 $\mu$s | 506 G, 300 K | Ref. [637] |
| | | N | 25.1 ms | 403 G, 3.7 K | Ref. [628] |
| | | $^{13}$C | 17.2 ms | | |
| | | $^{13}$C-$^{13}$C pair | 1.9 min | | Ref. [599] |
| | $T_2$ (echo) | e | 1.8 ms | 690 G, 300 K | Ref. [621] |
| | | N | 2.3 s | 403 G, 3.7 K | Ref. [628] |
| | | $^{13}$C | 770 ms | | |
| | $T_1$ | e | $> 1$ h | 403 G, 3.7 K | Ref. [598] |
| | | $^{13}$C | $> 6$ min | | Ref. [628] |
| **Gate time** | Single-qubit | e | $< 10$ ns | 850 G, 300 K | Ref. [618] |
| | Two-qubit | e-N | 700 ns | 513 G, 300 K | Ref. [617] |
| **Gate fidelity** | Single-qubit | e | 99.995% | 513 G, 300 K | Ref. [617] |
| | Two-qubit | e-N | 99.2% | | |

TABLE IV. Reported values on the NV center quantum platform.

Grover's search algorithm with very high efficiency [648], quantum-enhanced machine learning [649], and resonant quantum principal component analysis [650].

*Quantum network.*— With the help of flying photons, two remote NV nodes ($>$ 1 km) with memory qubits can be entangled [605, 651, 652]. Combining entanglement distillation [653] and deterministic entanglement delivery with more experimentally-favorable single-photon scheme [654], a three-node quantum network with more than one long-lived memory qubits has been realized [615, 655].

*Quantum sensing.*— Owing to the robustness and micro- or nanoscale features of diamond, NV centers have demonstrated the potential for high-sensitivity magnetic sensing in condensed-matter physics [656–661] and biophysics [662–664]. Moreover, NV sensors are also capable of detecting electric field [665–668], temperature [669, 670] and pressure [671]. Recently, the boundaries of NV quantum sensing have been pushed into some special or extreme-condition areas, such as at zero or low magnetic field [619, 672, 673], under high pressure [616, 674–678], and at high temperatures [679]. In parallel, quantum optimization algorithms [680] and QEC [681–683] have also been incorporated into quantum metrology so as to improve the sensitivity in the presence of noise.

*Outlook.*— Scalability is an inescapable question to which every candidate physical system for quantum computers should give an answer. There are two main challenges for the NV center-based quantum computing, namely device fabrications and multi-qubit control techniques. On the one hand, the deterministic schemes yielding NV centers with satisfactory precision are desired to produce large-scale functional devices [684, 685]. Meanwhile, since the underlying destruction and noises induced by the implantation may affect the coherence of the spin qubits [686], a trade-off solution for controllable production of NVs while preserving the coherence time should be developed. Accordingly, the demands in the fabrication arrays of nanostructures, such as nanopillars [687] and parabolic reflectors [688], which significantly enhance the collection efficiency, are also stringent. Additionally, it is also inspiring to explore the photocurrent-based electric readout of NV signals [689, 690]. On the other hand, with the growth of qubit number, the mechanism of noises in the system becomes more and more complicated. There is a need for techniques that integrate high-precision multi-qubit control techniques with decoupling techniques that suppress errors and crosstalk between multiple qubits.

## VII. NMR SYSTEM

*Introduction.*— NMR spectroscopy is a powerful and widely used analytical tool for the structural characterization of various organic matter. For nearly eighty years, it has spawned numerous scientific and technological applications in diverse areas of physics, chemistry, and life science. At the end of the twentieth century, motivated by a strong interest in quantum information science, there arose the idea of using liquid-state NMR to construct a quantum computer [7–9]. It was found that NMR is capable of emulating many of the capabilities of quantum computers, including unitary evolution and coherent superpositions. Actually, NMR quantum computing soon became one of the most mature technologies for implementing quantum computation [691, 692]. For instance, as early as 2001, researchers at IBM reported the first successful implementation of Shor's algorithm on a 7-qubit liquid-state NMR quantum computer [693]. Based on its well-established experimental technologies, NMR has now achieved universal control of up to 12 qubits [694–696], and allows investigation of a wide range of quantum information processing tasks, such as quantum simulation, quantum control, quantum tomography, and quantum machine learning. In the following, we briefly introduce the basics of NMR quantum computation and its impressive achievements.

*Basic Principle.*— In order to physically realize quantum information, it is necessary to find ways of representing, manipulating, and coupling qubits to implement non-trivial quantum gates, prepare a useful initial state, and readout the answer.

*Qubit.*— NMR quantum computation uses spin-1/2 nuclei in molecules to encode qubits. Due to the Zeeman effect, a spin-1/2 placed in an external magnetic field has two possible orientations, spin up $|\uparrow\rangle$ and spin down $|\downarrow\rangle$, which naturally offers a two-level system, or a qubit. In choosing a sample to be a quantum register, one property must be satisfied, that is, the spins should be distinguishable in frequency to allow individual qubit addressability. For heteronuclear molecules, because different types of nuclei have different gyromagnetic ratios, they can be easily distinguished. In the case of homo-nuclear molecules, although the nuclei have the same Zeeman splitting, they may stay in different electron environments, and the resulting nuclear shielding effect could induce different amounts of frequency shifts. In reality, precession frequencies for the nuclear spins can vary substantially, so it is best to choose such nuclei to form the quantum register. Fig. 7 shows the schematic of the NMR spectrometer and the commonly used molecules to encode qubits.

*Initialization.*—Conventionally, quantum computation starts from a pure state with all qubits initialized to the computational basis vector $|0\rangle$. However, due to the low polarization of the NMR spin ensemble at room temperature, it is practically rather difficult to get a genuine pure state. Alternatively, one can use the concept of pseudo-pure state (PPS) as a substitute [7]. PPS is a mixture of the maximally mixed state and a pure state, thus having similar behavior to that pure state under quantum gates and quantum measurements. To prepare PPS from the thermal equilibrium state, it would be necessary to involve non-unitary operations, which can be realized by applying gradient field pulses or utilizing relaxation effects. Currently, there exist a number of methods for PPS preparation, such as spatial averaging [698], line selective [699], and labeled-PPS [694, 700], etc. Ref. [701] analyzed and compared the efficiencies of these methods based on the theory of optimal bounds on state transfer under quantum channels. Overall, PPS has proven to be a convenient and useful tool for small-scale NMR quantum computation, yet when the number of qubits grows, there is a significant scalability challenge, i.e., the achievable purity of PPS scales very unfavorably. Approaches that attempt to address the scalability issue include algorithmic cooling [702] and parahydrogen-induced polarization [703], which have demonstrated the ability to prepare NMR spin systems with purities even above the entanglement threshold.

*Operation.*—One-qubit gates are just rotations on the Bloch sphere, which can be easily implemented in NMR with soft radio-frequency pulses. Soft pulses are usually of predefined shapes, such as the Gaussian waveform [704]. They contain energy only within a limited frequency range, and thus can selectively excite those spins that locate in this range. Therefore, a natural way to implement a single-qubit gate is to use a resonant, rotating Gaussian pulse with sufficient selectivity. But one should be careful that, when going back to the lab frame, there may be some phase errors that must be compensated to get the genuine target gate [705]. In NMR, a two-qubit gate is realized by making use of the natural $J$-coupling between the nuclei. In the case of multi-qubit gates, since all the $J$-couplings between the spins are evolving, one has to design refocusing schemes that are composed of a sequence of $\pi$ pulses to effectively turn off the unwanted $J$-coupling terms [706]. The usual way to quantify the level of coherent control is the randomized benchmarking protocol. Using randomized benchmarking, an average error rate for one- and two-qubit gates of $4.7 \pm 0.3 \times 10^{-3}$ on a three-qubit system was reported [707]. Another work has used a unitary 2-design and twirling protocol to estimate the average fidelities of Clifford gates on a seven-qubit NMR processor, finding an average experimental fidelity of 55.1% [708]. NMR has also explored other types of quantum gate implementation such as geometric quantum computation [709, 710]. Finally, we remark that NMR also provides non-unitary control means such as gradient field pulse and phase cycling, which are modeled as random unitary channels and can be used to destroy unwanted coherences.

*Measurement.*—NMR measurement is implemented by observing the free induction decay (FID) of the transversal magnetization of the spins by a detection coil wound around the sample. The recorded time-domain FID signal is Fourier transformed to obtain a frequency-domain spectrum, which is then fitted to obtain information about the spin state. Unlike projective measurements in other quantum systems, we can directly measure the expectation value of a single coherent Pauli observable in NMR, which is, in fact, an ensemble system. In order to measure the Pauli operators other than the directly observable single quantum coherences, it is necessary to apply an appropriate readout pulse to the spin state before acquiring the FID signal. To estimate an unknown quantum state, that is, to perform quantum state tomography, one needs to measure a complete set of basis operators. However, this is generally a challenging task since the number of degrees of freedom to be determined grows exponentially with system size.

*NMR quantum control.*—Over 50 years of development, researchers have developed abundant pulse control techniques in the NMR spin system, such as frequency selective pulses, composite pulses, refocusing schemes, and multiple pulse sequences, to name a few. These pulse techniques originated in demand for precise spectroscopy of complex molecules, and continue to be useful for NMR quantum information processing experiments [706]. Despite this, it is still desirable to further improve NMR control techniques to realize sufficiently high-fidelity gates that fulfill the fault-tolerant quantum computation requirement. Therefore, the interdisciplinary field of NMR and quantum control theory naturally arose, resulting in novel and more efficient pulse design and optimization techniques. For small-sized systems, one can employ time optimal control theory to reduce gate time so as to reduce decoherence effects [711, 712]. For relatively larger systems, it is usually hard to derive analytical control solutions, and then one needs to resort to numerical means. One of the most successful approaches in this regard is the GRAPE technique developed in Ref. [713], which is flexible, easy to use, and can produce smooth, optimal, and robust shaped pulses. GRAPE and its many variants have found broad applications not just in NMR but also in other experimental platforms. However, these numerical approaches are intrinsically unscalable. It would be a rather resource-consuming task to simulate controlled quantum evolution with a classical computer, even for a system with over tens of qubits. One possible approach to overcome this problem is to use subsystem-based quantum optimal control [705]. Another promising strategy is the hybrid-classical version of GRAPE, which employs a quantum simulator to efficiently simulate the controlled evolution [714]. This is essentially a closed-loop strategy, and has been experimentally tested first on a seven-spin system [714] and later on a twelve-spin system [696] to create multiple-correlated spin states. Besides scalability, noise is another major obstacle for high-fidelity quantum control, which could be addressed by robust control or open quantum system control. For example, more advanced DD sequences were put forward on solid-state NMR, resulting in much-improved robustness against different types of experimental errors while retaining good decoupling efficiency [715, 716]. It is worth mentioning that, the above mentioned control methods, such as composite pulse, GRAPE, spin echo, and DD, though developed from NMR firstly, are not at all restricted to NMR. Actually, many of these methods have already been successfully applied to other physical systems. Therefore, it is fair to say that NMR is an excellent platform and testbed for developing quantum control methods [705].

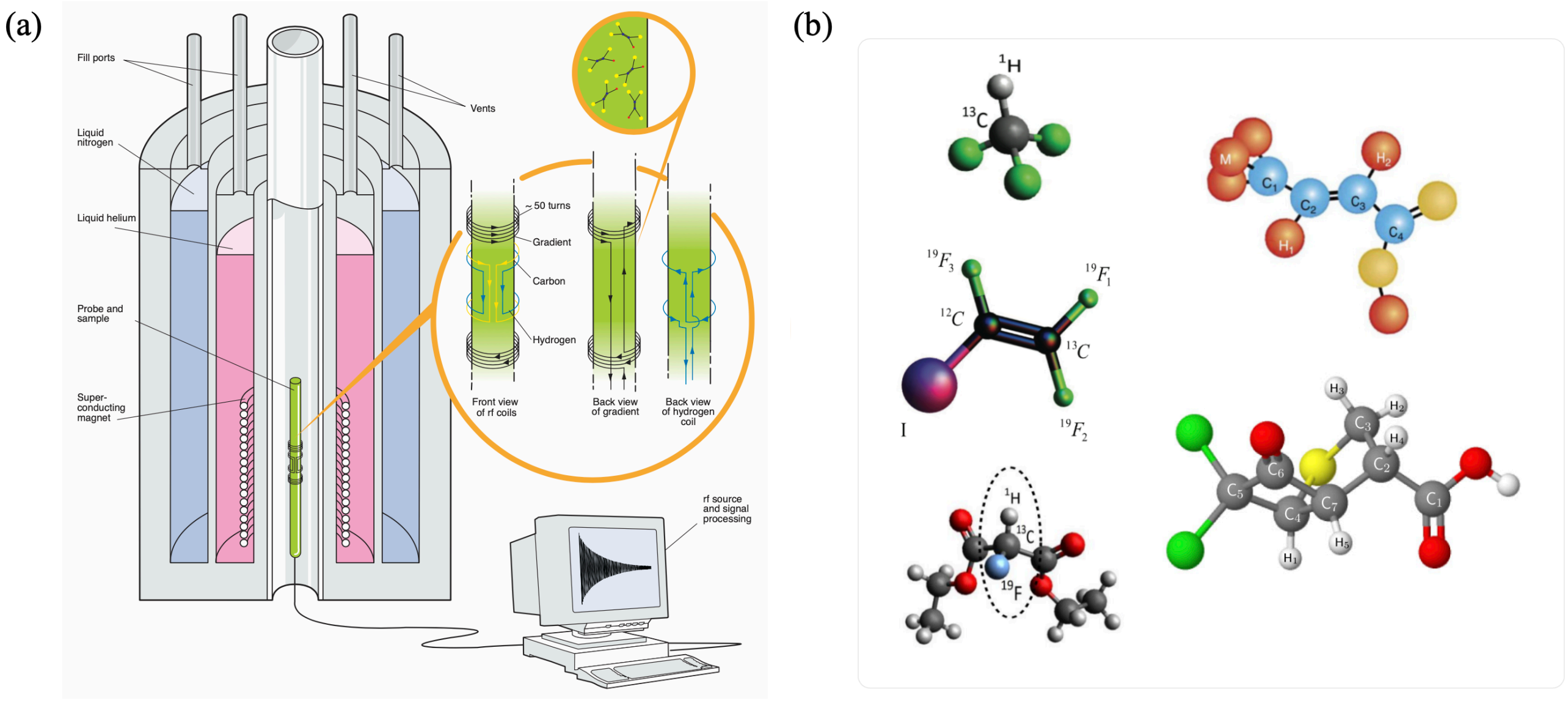

FIG. 7. (a) Schematic of a high-field liquid-state NMR spectrometer [697], which can be used for quantum computation. (b) A list of some commonly used molecules for nuclear spin quantum registers, ranging from two-qubit to twelve-qubit samples. The labels indicate nuclei $^{13}$C, $^{1}$H, or $^{19}$F (all having spin number 1/2) that are chosen as qubits candidates.

*NMR quantum processor.*—The NMR field has well-established quantum control methods and experimental technologies, enabling a series of influential fundamental or applicative researches in quantum computing, quantum simulation, quantum cloning [717–719], QEC [720, 721], quantum thermodynamics [722–725], quantum contextuality [726], etc. In the following, for short, we only review a few developments related to quantum algorithms, quantum simulation, and quantum learning.

*Quantum algorithm.*—Since the early stage of NMR-based quantum computing, there have been reported experimental realizations of some of the well-known quantum algorithms, such as Deutsch-Jozsa algorithm [727] on a two-qubit carbon-13 labeled chloroform molecule, Grover's search algorithm [728] on another two-qubit sample partially deuterated cytosine, QFT algorithm on a three-qubit sample [729], and Shor's quantum factoring algorithm [693] on a seven-qubit system.

*Quantum simulation.*—NMR has been used as a quantum simulator to explore a variety of interesting quantum phenomena, ranging from quantum many-body physics, quantum chemistry, biology, and even cosmology. Simulating the equilibrium and non-equilibrium dynamics of quantum many-body systems is one of the most fascinating topics in the field of quantum simulation, and NMR seems to be very suitable for this task. For example, a three-spin frustrated magnet was simulated with NMR, in which the phase of the system as a function of the magnetic field and temperature was explored [730]. The phase diagram of the ground state of the Hamiltonian with three-body interactions was simulated [731] and the phase transition of the long-range coupling model was first observed by monitoring Lee-Yang zeros [732]. In another research, the authors employed a four-qubit NMR simulator to explore the use of out-of-time order correlators to probe quantum information scrambling [733] and equilibrium or dynamical quantum phase transitions [734] in a chaotic Ising chain model. NMR has also found applications in various chemistry problems by directly simulating molecules or chemical reactions, such as computing the ground-state energy of a hydrogen molecule [735], finding the energy spectrum of a water molecule [736], and exploring the prototype laser-driven isomerization chemical reaction dynamics[737]. Besides, the NMR quantum simulator can also be used to investigate topological orders by simulating the ground state of a topological Hamiltonian [738–741].

*Quantum machine learning.*—NMR has been one of the experimental platforms where quantum machine learning algorithms can be demonstrated, and it is in the initial stages of exploring the use of quantum machine learning to directly process classical image information. For instance, a hand-written image recognition task to discriminate between 9 and 6 is realized by implementing a quantum support vector machine on a four-qubit NMR processor [742]. The boundary that separates different regions of an image is detected experimentally by implementing a quantum image processing algorithm [743]. Quantum principal component analysis, an important tool for pre-processing data in machine learning, has also been experimentally implemented on NMR for the first time for small-scale human face recognition tasks [744].

*Outlook.*—Primary challenges for liquid-state NMR quantum computation include a lack of appropriate molecules to serve as quantum registers, unavailability of high-purity quantum states and quantum resources such as entanglement, and difficulty in achieving scalable and high-fidelity control on large spin systems. One approach that may overcome some of these limitations is to shift to solid-state NMR. Solid-state NMR has already been used in demonstrating quantum heat engine [745], exploring many-body localization [746, 747], and observing prethermalization [748, 749]. Some other promising approaches that are closely related to NMR are the silicon-based nuclear spin quantum computer, which is a hybrid between the quantum dot and the NMR [10], and the recent technology of nuclear electric resonance [750]. Finally, while NMR has certain intrinsic difficulty in becoming a scalable route to large-scale quantum computation, the many lessons learned in the past decades' research are very likely to be relevant for advancing the development of other quantum technologies.

## VIII. NEUTRAL ATOM ARRAYS

Over the past two decades, deterministically prepared neutral atom arrays have emerged as a promising platform for quantum computing and quantum simulation [751–755]. Controlled interactions between atomic qubits are mediated by the long-range dipole-dipole interactions via Rydberg states. These long-range Rydberg interactions allow creating specific quantum Hamiltonians and easy analog quantum simulations. They are also the workhorse for constructing digital gates and realizing any physical models. Most experiments to date focus on alkali atoms Rb and Cs, which have single valence electrons and can be simply laser-cooled and manipulated. In recent years, with two valence electrons, alkaline-earth(-like) elements Sr and Yb have attracted growing attention due to their appealing features, such as narrow and ultra-narrow optical transitions and magic-wavelength optical traps for Rydberg states [756–759]. The schematic diagram of atomic qubits and their pros and cons are summarized in Fig. 2(e).

*Scalability.*—A neutral atom quantum computer is based on an array of single atoms localized in optical tweezers, as depicted in Fig. 8. Quantum information is encoded in electronic spin states of alkali atoms or nuclear spin states of alkaline-earth(-like) atoms. Neutral atom platforms have a notable advantage for scalability. It is relatively easy to expand the number of atomic qubits. In a microscopic optical tweezer, either one atom or zero atoms can be trapped each with a probability of roughly 50% due to light-assisted collisions [760, 761]. The stochastic loading efficiencies have been enhanced to 80%–90% for alkali species and 96% for alkaline-earth(-like) species by accurately tuning parameters under blue-detuning lasers and using artful cooling techniques [762–766]. After probabilistic loading in tweezers, single atoms can be rearranged into defect-free arbitrary patterns using a real-time control system and dynamically moving tweezers [767–769]. Up to date, large-scale platforms consisting of more than 100 neutral atoms have been created, such as an array with an average number of 110 $^{133}$Cs atoms [770], defect-free square and triangular arrays of 196 and 147 $^{87}$Rb atoms [771], and a defect-free programmable array of up to 256 $^{87}$Rb atoms [772]. The rearrangement process takes a total time of hundreds of milliseconds and results in a high filling fraction of up to 99% [771, 772]. In the rearrangement of larger arrays, atom losses from tweezers will limit filling fractions, as the rearrangement time will increase with the system size. The typical trap lifetime is about tens of seconds due to collisions with background gas in a vacuum chamber. In order to reduce the residual gas pressure, optical tweezers can be placed in a cryogenic environment at a temperature of a few kelvins. The trapping of single Rb atoms in cryogenic arrays of optical tweezers has been demonstrated with a measured lifetime up to 6000 s, a 300-fold improvement compared to the room-temperature setup [773]. In this cryogenic experimental setup, large arrays consisting of more than 300 $^{87}$Rb atoms have been realized with an unprecedented probability of $\sim$ 37% to prepare defect-free arrays [774]. We anticipate that the number of atomic qubits in a neutral atom processor will be increased from hundreds to thousands. Furthermore, several individual processors will be coupled together with atom-photon interconnects.

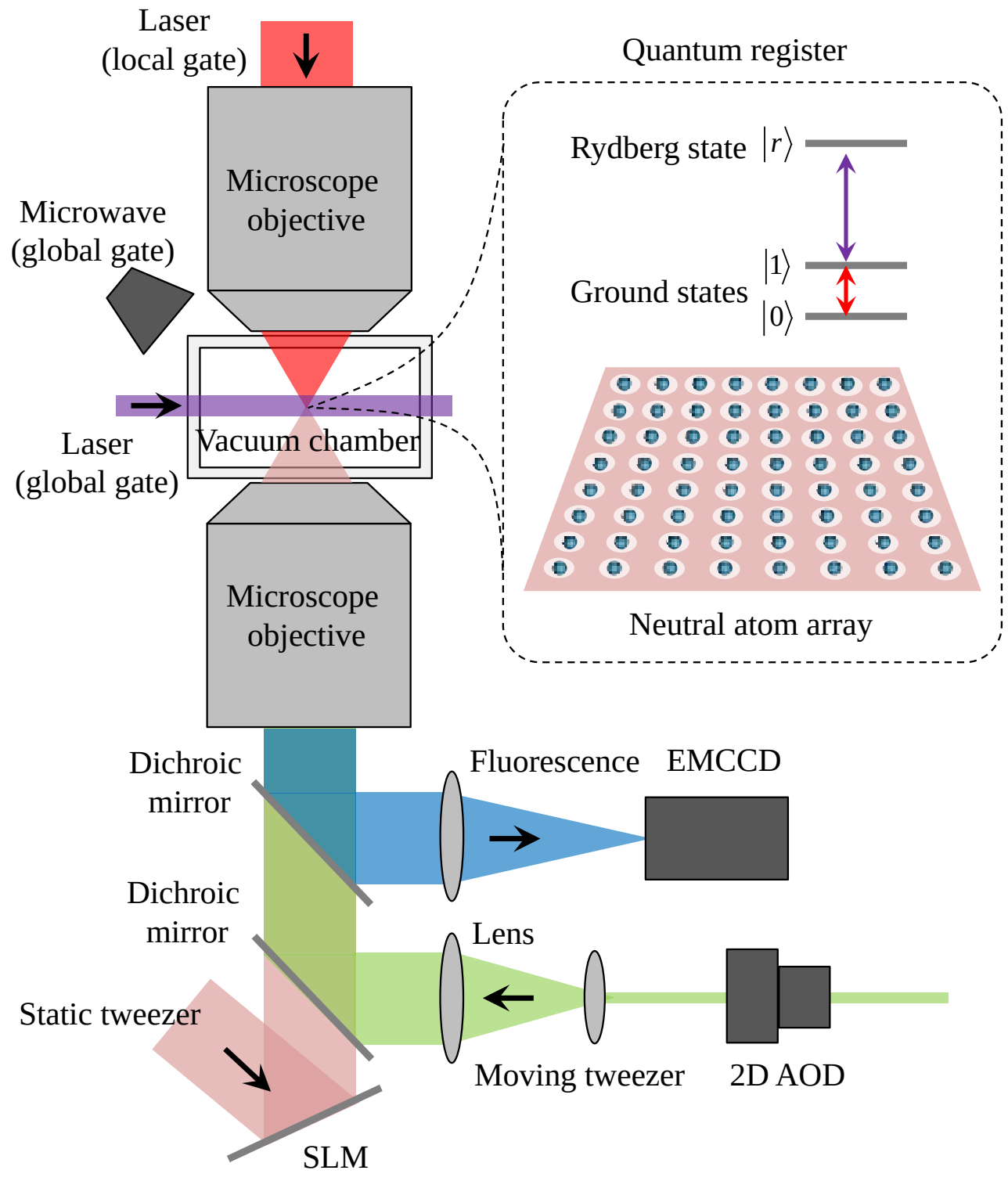

FIG. 8. Schematic of a neutral atom quantum computer. In a microscopic tweezer array, single atoms are rearranged into defect-free arbitrary patterns. Atomic qubits can be encoded in electronic spin states or nuclear spin states. Single-qubit operations are performed through microwave or optical spectroscopy. Two-qubit gates and entanglement are realized based on long-range Rydberg interactions. EMCCD, Electron Multiplying Charge-Coupled Device. SLM, Spatial Light Modulator. 2D AOD, Two-Dimensional Acousto-Optic Deflector.

*Initialization and readout.*—Atomic qubits encoded in the internal states can be initialized using optical pumping techniques. The estimated preparation fidelity for single alkalis is $F > 99.5\%$ [775]. A widely used technique for qubit readout relies on the state-selective ejection of neutral atoms. When illuminating with a resonant laser pulse, any atoms in one state are pushed out of the optical tweezers, whereas atoms in the other state are not influenced and remain trapped. Subsequently, trapped atoms are detected by collecting laser-induced fluorescence which is not state-selective. The typical measurement fidelity is $F > 98\%$ [775]. Atom losses in this technique prevent from measuring qubits in the middle of quantum circuit execution. As an alternative, for a lossless fluorescent state detection, only a small number of photons should be scattered to minimize the atom heating. This approach has been demonstrated for one qubit and multiple qubits in optical tweezers [776–779]. In the future, atom losses due to heating will be reduced to a level that allows implementing repetitive QEC for quantum computation.

*Gates.*—We summarize the state-of-the-art gates implemented in Table V. Single-qubit operations are performed through microwave or optical spectroscopy. In large atom arrays, a universal approach for single-site addressing is the focused laser beam scanning or the application of static field gradients. The gate operation time is $t_1 = 0.1 - 10$ $\mu$s. Single-qubit gate errors are caused by the fluctuations of the pulse amplitude and detuning [775, 780]. An average fidelity of $F_1 = 99.83\%$ was measured in the $^{133}$Cs experiment using randomized benchmarking [781]. Recently, the gate fidelities for $^{87}$Rb atoms have been enhanced by using composite pulse sequences, which make gate errors highly insensitive to pulse errors [782]. The estimated gate errors are about $3 \times 10^{-4}$. In the platform of alkaline-earth(-like) atoms, the average single-qubit gate error of $5.2 \times 10^{-3}$ has been extracted [766].

Controlled interactions between neutral atoms are a fundamental requirement for entangling particles. One strategy is to implement local entangling operations via ultracold spin-exchange interactions, which has been demonstrated with two individual atoms in movable tweezers [783]. However, it is a great challenge to overlap the atomic wavefunctions in neutral atom arrays. Another strategy is based on the strong and long-range dipole-dipole interactions between Rydberg atoms. When two atoms are in close proximity to each other, two-qubit gates and entanglement have been realized via the Rydberg blockade effect [784, 785]. Most researchers give attention to the latter one owing to its feasibility. The corresponding gate time is $t_2 = 0.4$–$2$ $\mu$s. For two-qubit gates via Rydberg interactions, the dominant sources of gate errors are the ground-Rydberg Doppler dephasing, the spontaneous emission from the intermediate state in the two-photon excitation process, and the excitation laser phase noise [775]. In the $^{87}$Rb atom arrays, the fidelities of the two-qubit entanglement operations have been extracted to be $F_2 \geq 97.4\%$ by sup-pressing Rydberg laser phase noise via a reference cavity [786, 787]. In an array of $^{171}$Yb atoms, a two-qubit gate with the fidelity of $F_2 = 83\%$ has been firstly demonstrated in [788]. In this experiment, the gate error is attributed to Raman scattering from the gate beam and autoionization from a small Rydberg population. For two individually trapped $^{88}$Sr atoms, a Bell state has been created with a high fidelity of $> 99.5\%$, in which qubits are encoded in a metastable state and a Rydberg state [789].

In the Rydberg excitation process, we must consider the effect of the different trapping potentials for both ground and Rydberg levels. In the experiments with single $^{87}$Rb or $^{133}$Cs atoms, the tweezers are turned off for a short duration to mitigate anti-trapping of the Rydberg states. Atom losses and heating limit the Rydberg excitation time and the number of excitation loops. When using 0.5 $\mu$s drops for each two-qubit gate, hundreds of drops can be made before atom loss becomes significant [782]. It should be noted that the ion core polarizability of the alkaline-earth(-like) atoms can be used to trap Rydberg states in conventional, red-detuned optical tweezers. The Rydberg states of single $^{174}$Yb atoms have been stably trapped by the same red-detuned optical tweezer that also confines the ground state [790]. Therefore, the interaction time of Rydberg states for alkaline-earth(-like) atoms can be extended.

*Coherence.*—Neutral atoms are well isolated from the environment and exhibit long coherence times. For Rb and Cs atoms [782, 791], the hyperfine qubit relaxation time is about $T_1 \sim 4$ s, which is limited by the spontaneous Raman scattering of photons from the trapping laser. The inhomogeneous dephasing originates from the energy distribution of trapped atoms. The typical dephasing time is $T_2^* \sim 4$ ms. For the homogeneous dephasing, common mechanisms are the intensity fluctuations of the trapping laser, magnetic field fluctuations, and heating of atoms. The homogeneous dephasing time of $T_2' \sim 1$ s has been observed using XY8 and XY16 DD sequences [716, 792]. By analyzing these mechanisms, we observe that the differential light shift of qubit states is the root of the dephasing. A magic-intensity trapping technique allows mitigating the differential light shift. The coherence time has been enhanced to 225 ms, where the extracted inhomogeneous dephasing time is $T_2^* \sim 1.5$ s [793]. In comparison with the electronic spin qubits, nuclear spin qubits in alkaline-earth(-like) atoms are robust to perturbation by the optical tweezers. The estimated coherence time of single $^{87}$Sr atoms are $T_1 \gg 10$ s, $T_2^* = 21$ s, and $T_2' = 40$ s in the spin echo process [794]. In addition, coherence properties of single $^{171}$Yb atoms have been measured [766, 788], as listed in Table V.

*Digital quantum operations.*—Digital gate-based circuits on programmable neutral atom processors were demonstrated by two experimental groups. In Ref. [791], researchers at Wisconsin employed an architecture based on individual addressing of single atoms with tightly focused beams. Quantum circuits were decomposed into global microwave rotations, local phase rotations and local two-qubit CZ gates. Scanning Rydberg excitation beams enabled coherent and simultaneous addressing of pairs of atoms. In this platform, researchers demonstrated the preparation of GHZ states with up to 6 qubits, quantum phase estimation algorithm for a chemistry problem, and QAOA for the maximum cut graph problem. In Ref. [782], researchers at Harvard employed another architecture based on the coherent transport of entangled neutral atoms. Two-qubit CZ gates were implemented in parallel by two global Rydberg laser beams. Subsequently, entangled qubits were coherently transported to change the connectivity and perform the next layer of quantum operations. This architecture was used to generate a 12-qubit cluster state, a 7-qubit Steane code state, topological surface, and toric code states. Finally, researchers realized a hybrid analogue-digital evolution and measured the entanglement entropy. These results represent a key step toward realizing a quantum computer with neutral atoms.

*Analog quantum operations.*—Combined with the wide tunability of array geometry, Rydberg atom arrays are suitable to implement various Hamiltonians [753–755]. When the spin states are encoded in the ground level and the Rydberg level, the quantum Ising-like models are obtained. In a one-dimensional chain with up to 30 atoms and a $7 \times 7$ atoms array, the excitation dynamics and the pair correlation functions of quantum Ising models were explored after suddenly switching on the Rydberg excitation pulse [796, 797]. The similar quench dynamics were also studied in the linear and zigzag chains [798]. For three-dimensional arrangements of Rydberg atoms, quantum Ising Hamiltonians mapped on various connected graphs were constructed with tens of spins [799, 800]. Sweeping the Rydberg excitation detunings allows probing more abundant many-body dynamics. Quantum phase transitions into $\mathbb{Z}_n$ ordered phases and the critical dynamics were demonstrated in a one-dimensional chain with tunable interactions [801, 802]. Antiferromagnetically ordered states were further explored in two-dimensional arrays with up to hundreds of atoms [771, 772, 803]. Although the Rydberg interactions generally lead to thermalization in many-body systems, it was realized that quantum many-body scars avoided rapid thermalization when preparing the two-dimensional atoms array in the antiferromagnetic initial state [804]. Besides the Ising-like models mentioned above, recent works include observing topological phases in a quantum dimer model and a Su-Schrieffer-Heeger model [805, 806], engineering the XXZ spin model using a periodic external microwave field [807], and investigating quantum optimization algorithms for solving the maximum independent set problem [795].

*Outlook.*—Recent breakthroughs in Rydberg atom arrays exhibit the ability to study many-body physics and realize highly programmable and scalable quantum com-puting. Primary challenges for this platform are higher fidelity of two-qubit gates, quantum nondemolition measurements, and low crosstalk between individual qubits in a large array. Two-qubit gate fidelity can be improved by further cooling alkali atoms via Raman sideband cooling [808, 809] or using alkaline-earth(-like) elements that have exhibited excellent gate performance. Combined with lossless fluorescent state detection, introducing a second atomic element allows monitoring quantum processors via quantum nondemolition couplings to auxiliary qubits [810]. In addition, a dual-element platform enables low crosstalk manipulations of the homonuclear and heteronuclear interactions when increasing system size [811].

TABLE V. Reported state-of-the-art performance of neutral-atom qubits. $T_1$ is the spin relaxation time. $T_2^*$ and $T_2'$ refer to the inhomogeneous and homogeneous dephasing time. $F_1(F_2)$ and $t_1(t_2)$ are the gate fidelity and the operation time of single(two)-qubit manipulation. $N_\mathrm{d}$ and $N_\mathrm{a}$ refer to qubit numbers in digital quantum processors and analog quantum simulators, respectively.

| **Property** | | **Alkali atom** | **Alkaline-earth(-like) atom** | |
|---|---|---|---|---|
| | | **Electronic spin** | **Nuclear spin** | |
| | | **$^{85,87}$Rb/$^{133}$Cs** | **$^{87}$Sr** | **$^{171}$Yb** |
| **Coherence** | $T_1$ | 4 s [782, 791] | $\gg 10$ s [794] | $10-100$ s [766] |
| | $T_2^*$ | 4 ms [782, 791] | 21 s [794] | 3.7 s [766] |
| | $T_2'$ | $\sim 1$ s[a][782, 791] | 40 s[b][794] | 7.9 s[b][766] |
| **Gate time** | $t_1$ | $0.1-10$ $\mu$s | | 0.7 $\mu$s [766] |
| | $t_2$ | $0.4-2$ $\mu$s | | 0.9 $\mu$s [788] |
| **Gate fidelity** | $F_1$ | $\sim 99.97\%$[c][782] | | 99.48%[d][766] |
| | $F_2$ | 97.4%[e][787] | | 83%[e][788] |
| **Qubit number** | $N_\mathrm{d}$ | 6[f][791], 24[g][782] | | |
| | $N_\mathrm{a}$ | 289 [795] | | |
| **Environment** | | Ultra-high vacuum $P\sim 10^{-11}$ Torr, magnetic field $B\sim 10$ G | | |

[a] XY8 and XY16 dynamical decoupling sequences.
[b] Spin echo process.
[c] This is estimated from the scattering limit which is consistent with an accumulated error. Randomized benchmarking will be applied in the future.
[d] Randomized benchmarking.
[e] Bell state fidelity.
[f] GHZ state based on individual addressing of single atoms.
[g] Toric code state based on coherent transport of entangled atoms.

## IX. PHOTONIC QUANTUM COMPUTING

*Introduction.*—Amongst all platforms for quantum computing, photon has several unparalleled advantages: i) one of the best candidates for room-temperature quantum computing, owing to the inherent advantage that it weakly couples with the surrounding environment, and ii) natural interface for distributed quantum computing, which acts as a flying qubit to connect many quantum nodes, and iii) compatible with CMOS technologies, bringing optical quantum computing to a new cutting-edge stage.

Optical quantum computing can be dated back to 2001, when Knill, Laflamme and Milburn (KLM) pointed out that it is possible to create universal quantum computing solely with linear optical elements [812]. This landmark work opens a way for linear optical quantum computing. However, the daunting resource overhead makes the KLM scheme extremely hard to implement. In 2010, a much more feasible and intermediate model—boson sampling—was proposed and analyzed by Aaronson and Arkhipov [813]. Compared to the KLM scheme, boson sampling is a much easier linear optical quantum computing model which can beat all classical computers with only 50–100 photons, but at a cost that it's no longer universal. In 2017, a variant called Gaussian boson sampling (GBS) was developed by Hamilton *et al.*, in which the input is replaced as single-mode squeezed states rather than single photons [814]. It is a new paradigm that not only can provide a highly efficient approach to large-scale implementations but also can offer potential applications in graph-based problems and quantum chemistry.

In the past two decades, we have witnessed great progress in linear optical quantum computing [815, 816], especially on single-photon sources, linear optical networks, and single-photon detectors. These achievements have enabled a series of essential experimental results in the preparation of large-scale entangled states [817–819], and quantum computational advantage through boson sampling [16, 18, 820].

*Photon qubit.*— Photon has the richest degrees of free-dom to encode as a qubit. In the following, we summarize several frequently-used bases for photonic qubits.

• *Polarization*: The qubit can be encoded on the two orthogonal geometrical orientations of electromagnetic field. It is widely used in linear optical quantum computing.

• *Path*: Two transmission paths of single photons can be a qubit. Usually, the phase of the path in free space is unstable, while it's perfect for integrated photonics.

• *Time bin*: The former and latter arrival time of single photons is encoded as a qubit.

• *Frequency bin*: Frequency-bin qubit refers to the superposition state of a photon with two different frequencies (colors).

• *Photon number*: Vaccum and single-photon states encode qubit's 0 and 1, respectively.

• *Orbital angular momentum (OAM)*: OAM describes the spatial distribution of light. In quantum theory, OAM of a photon has a value of $L_z = m\hbar$. Any two OAM states with different $m$ values form a photonic qubit.

*Quantum light sources.*—Single-photon source is one kind of quantum light source that emits one and only one photon at a certain time, in a well-defined polarization and spatial-temporal mode. Specifically, the single photons should possess the same polarization, spatial-temporal mode and transform-limited spectral profile for a high-visibility Hong-Ou-Mandel-type quantum interference [821].

Spontaneous parametric down-conversion (SPDC) sources [822, 823] play a vital role in many fundamental quantum optics experiments. Notably, this year's physics Nobel prize is partially "for experiments with entangled photons". However, SPDC is intrinsically probabilistic and unavoidably mixed with multiphoton components. The single-photon efficiency is typically as low as ∼1% to suppress unwanted two-photon emission. To overcome this issue, an alternative way is to multiplex many SPDC sources to boost the efficiency of single-photon sources [824]. Another approach is directly generating high-quality single photons from a two-level system. Amongst all platforms [825–832], semiconductor quantum dots [833] provide state-of-the-art single-photon sources with an overall efficiency of 57% [834]. This mainly benefits from a polarized microcavity developed by Wang *et al.* [835], which has a polarization-dependent Purcell enhancement of single-photon emission so that the overall efficiency can surpass the 50%. In near future, the single-photon efficiency can be improved over 70% by better sample growth and boosted collection efficiency, which should surpass the efficiency required for universal quantum computing [836].

In quantum optics, another quantum light source is squeezed state, which refers to a quantum state that the uncertainty of the electric field strength for some phases is smaller than that of a coherent state. Such a state is commonly generated by strongly pumping nonlinear mediums [837]. It was shown that continuous-variable (CV) quantum computing can be constructed [838] by using squeezed states and simple linear optical elements, such as beam splitters and phase shifters. Then, Gottesman, Kitaev, and Preskill (GKP) proposed a robust QEC code over CVs to protect against diffusive errors [839]. Until now, the record squeezing is 15 dB from type I optical parametric amplifier [840], and many quantum experiments were executed towards large-scale CV quantum computing [841–843]. Compared to discrete-variable (DV) quantum computing, CV has a valuable feature that the entanglement can deterministically emerge by mixing two squeezed states with a simple beam splitter, while it is hard to obtain in DV case since such a nonlinear interaction is so weak that we have to generate entanglement by a conditional fashion, namely as post selection. Nevertheless, CV quantum information processing can never be perfect, because the quality of entanglement strongly depends on the amount of squeezing that it's extremely sensitive to the loss.

*Linear optical networks.*—The interferometer acts as a unitary transformation on the single-photon Fock state or the single-mode squeezed state. In 1994, Reck *et al.* showed that a universal unitary transformation can be realized by beam splitters and phase shifters arranged in a triangular configuration [844]. In this scheme, the optical depth is $2(n-1)-1$, the number of beam splitters is $\frac{n(n-1)}{2}$, where $n$ is the number of modes. In 2016, Clements *et al.* demonstrated that an interferometer with a rectangular configuration is equivalent to a triangular one [845]. Moreover, the optical depth is reduced to $n-1$, and the number of beam splitters is reduced to $\frac{n^2-2n+2}{2}$. This is a more compact and robust design with a symmetry configuration.

For boson sampling, the linear optical network should combine high transmission, Haar randomness, high spatial and temporal overlap simultaneously. There are many different implementation approaches, such as micro-optics [16, 820, 846, 847], time-bin loops [18, 848], and integrated photonic circuits [849–851]. For universal quantum computing, it should further be programmable. Micro-optics possesses the highest transmission efficiency, while it lacks the demonstration of programmability. Time-bin loops and integrated on-chip circuits are programmable but suffer from serious losses. How to reduce the losses meanwhile programmable is a long-sought goal in the future.

*Boson sampling.*—In 2011, Aaronson and Arkihpov argued that a passive linear optics interferometer with single-photon state inputs cannot be efficiently simulated [813]. This model is so-called boson sampling, a non-universal quantum computing model much easier to build than universal quantum computing. In boson sampling, $n$ identical bosons are sent into an $m$-mode ($m \gg n$) Haar-random interferometer and sampling the output distribution in the photon number basis. Because of the bosonic statistics, the probability amplitudes of the final state are related to the permanent of submatrices, a problem known to be in the complexity class of #P-complete. It is strongly believed that a moderate-size boson sampling machine, even an approximate one with a multiplicative error, will be intractable to be simulated with state-of-the-art classical computers [813]. More importantly. boson sampling is a strong candidate to demonstrate quantum computational supremacy [852], an important milestone in the quantum computing field.

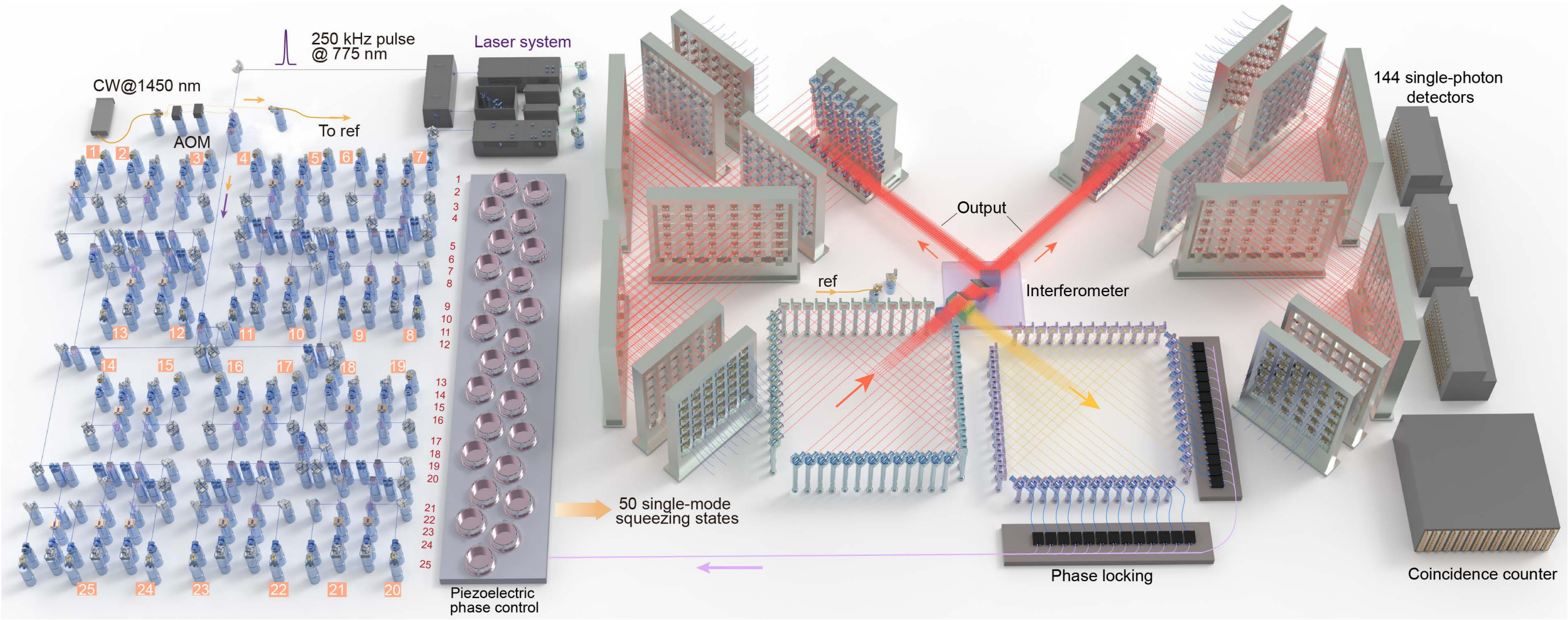

FIG. 9. Experimental setup of *Jiuzhang*. It is mainly composed of five parts. At the upper left region, high-intensity transform-limited pulse laser with wavelength of 775 nm are prepared to pump 25 two-mode squeezed state (TMSS) sources (at the left region, labeled in orange). Meanwhile, continuous-wave 1450-nm laser are guided and co-propagates with the 25 TMSS sources. The 1550-nm two-mode squeezed light is collected into temperature-insensitive single-mode fiber, of which 5-m bare fiber is winded around a piezo-electric cylinder to control the source phase (at the center region). In the center-right region, by using optical collimators and mirrors, 25 TMSSs are injected into a photonic network and 25 corresponding light beams (colored in yellow) with wavelength of 1450 nm and intensity power of about 0.5 $\mu$W are collected for phase locking. The 144 output modes are distributed into four parts using arrays of tunable periscopes and mirrors. Finally, the output modes are detected by 144 superconducting nanowire single-photon detectors and registered by a 144-channel ultra-fast electronics coincidence unit.

In 2013, simultaneously, four groups reported the small-scale proof-of-principle boson sampling experiments [849–851, 853]. Indeed, the final photon distribution is proportional to the square of permanent modulus. However, all these experiments are based on SPDC sources, which were intrinsically probabilistic and mixed with multi-photon components. In an attempt to solve the intrinsic probabilistic problem of SPDC, scattershot boson sampling was proposed in 2014 by Lund *et al.* [854] and was first demonstrated in 2015 by Bentivegna *et al.* [855]. Then, Zhong *et al.* improved the photon number to five [817]. Though scattershot Boson sampling is theoretically beautiful, it is hard to realize quantum advantage experimentally due to the extensive challenges of ultra-high heralding efficiency, fast optical switches, and an excess of SPDC sources required.

A direct way to solve the problems of SPDC is directly using on-demand single photon sources based on coherently driving a quantum two-level system. In 2017, Wang *et al.* successfully performed the first five-photon boson sampling experiment using an actively demultiplexed quantum-dot single-photon source and an ultra-low loss photonic circuit, and showed a high sampling rate that is 24,000 times faster than all previous experiments, beating early classical computers — ENIAC and TRADIC [846]. In 2019, Wang *et al.* demonstrated a boson sampling with 20 input photons and a 60-mode interferometer [847]. Finally, at most 14 photons are detected at the output, and the output state Hilbert space reaches up to $3.7\times10^{14}$ dimensions, which is over 10 orders of magnitude larger than the previous works.

A more efficient way to demonstrate quantum computational advantage is through GBS, thanks to the single-mode squeezed state inputs, of which more than one-photon components are allowed while its computational complexity is as hard as original boson sampling. In 2020, a landmark experiment was executed by Zhong *et al.* [16] and successfully demonstrated quantum computational advantage. Then, GBS was improved with 50 single-mode squeezed-state inputs and a 144-mode interferometer, and up to 113-click coincidences are detected [820] (see Fig. 9). These rudimentary photonic quantum computers, named as *Jiuzhang*, in honor of an ancient Chinese mathematical classic — "The Nine Chapters on the Mathematical Art", yield an output state space dimension of $10^{43}$ and a sampling rate that is $10^{10}$ faster than using the state-of-the-art simulation strategy and supercomputers.

Technically, *Jiuzhang* is partially programmable owing to the precisely controlling the phases of input TMSSs. To make boson sampling fully programmable, a notable way is to encode photonic modes to time bins [856]. In this approach, the splitting ratio and phase of every time bin can be changed as will in real time. In 2017, He *et al.* reported the first time-bin-encoded boson sampling combining a on-demand quantum dot single photon source [848]. It is worth noting that this time-bin multimode network is fully electrically programmable, and fully connected (without zero elemments). Recently, combining the aforementioned GBS approach, Madsen *et al.* reported quantum advantage by building a GBS machine with time-bin loops, where 216 single-mode squeezed state inputs and 16 photon-number-resolving detectors are used, and finally up to 219 photons are detected [18]. Noting that this machine is partially programmable, and most of the matrix elements are zeros, namely partially connected.

Because the output probability of GBS is related to hafnian, which corresponds to the number of perfect matching of a graph, it links to several potentially practical applications [857–860]. Next, GBS will naturally be developed as a special-purpose photonic platform to investigate these real-world applications, as a step toward NISQ processing [861].

*Scalability.*—For large-scale quantum information processing, the photonic platform faces two major hurdles in the current stage — the loss and the ultra-weak interaction between independent photons. In principle, the loss is both locatable and detectable, it should be much easier to be solved than computational errors (such as X, Z errors). For DV quantum computing, theoretical analysis [862, 863] suggests that at most 50% loss is allowed for scalable quantum computing, which is much less stringent and restrictive than a threshold of ∼1% for surface code to address computational errors. In one-way quantum computing, some works pointed out that for $m$-photon cluster state, at least a fusion success probability of $1/(m-1)$ is required for universal quantum computation [864], while the upper bond needs further works in the future. For instance, the upper percolation threshold of the 3-photon clusters is bonded by 0.5898 [864]. For CV quantum computing, loss will cause the squeezed states to move closer to the vacuum state, meanwhile losing their quantum feature. The exciting thing is that the fault-tolerant quantum computing with GKP qubits only requires a squeezing level of $\sim$ 10 dB [865], which allows a high loss threshold for scalable quantum computing. In summary, losses in both DV and CV photonic quantum computing can be handled with a relatively large threshold.

Photon-photon interaction at a single-photon level is a fundamental question both in photonic quantum computing and quantum optics. It is strongly believed that nonlinear interactions are needed to deterministically generate entanglement between photons [866]. Over the last two decades, several approaches were developed to address this issue, such as electromagnetically induced transparency [867], atom-cavity interaction [868], atom-atom interaction [869] and atoms in chiral waveguide [870], etc. In 2016, Hacker *et al.* reported a photon-photon gate with an efficiency of 4.8% and a fidelity of 76.2% [871], which suffers from inefficient photon storage and retrieval during the whole process, and the gate fidelity is limited by the precision of spin characterization. The same issues happen to several recent experiments utilizing Rydberg blockade [872–874]. By storing single photons in a long-lived Rydberg state, the efficiency of single-photon storage and retrieval has been improved to 39% [873]. In the future, by harnessing the strong nonlinearity in $\chi^{(2)}$ mediums and cQED systems, photon-photon gates can be realized with both high fidelity and efficiency which surpass the thresholds required for fault-tolerant quantum computing. In this case, the photonic platform will provide a perfect stage and a potential leading platform working at room temperature for fault-tolerant quantum computing.

## X. OUTLOOK AND CONCLUSION

We have reviewed prominent quantum computing platforms that have seen significant advances over the last decade. Currently, these quantum platforms are in different stages of maturity, where each system exhibits both advantages and limitations. To achieve better control fidelity and scalability of the various quantum platforms, challenges must be addressed to match the requirements for large-scale quantum computing on different platforms. For solid-state quantum systems, high-quality materials and advanced fabrication technologies are essential for the quality of qubits. For example, low-charge noise interfaces are critical for semiconductor qubits, which could be improved by importing industrial techniques of IC and encouraging a transfer from laboratory-level engineering to foundry-level fabrication. For photonic atom-based qubit systems, fundamentals and advanced techniques in precise control of individual atoms and atom-photon interconnectors between multiple processors could be developed. In the cases of NMR systems, one approach that may overcome difficulty in achieving scalable and high-fidelity control on large spin systems in liquid NMR quantum computation is to shift to solid-state NMR. While for qubits based on NV centers, the deterministic and controllable production of NV centers while preserving the coherence time needs to be further developed.

For all quantum computing platforms, the typical overheads for scalable quantum computing, such as crosstalk, gate heating, and frequency crowding, should also be considered carefully. The techniques that integrate high-fidelity control techniques, such as DD schemes that suppress errors and crosstalk between multiple qubits, are necessary for high-fidelity multi-qubit control with the growing number of qubits, where system size should increase without compromising control quality. By merging the techniques developed for quantum computing, we might also gain better performance in quantum metrology and quantum simulations. In addition to techno-logical advances to address the challenges, efforts in integrating multiple quantum platforms may also be essential for utilizing the advantages of different platforms for future applications. For instance, tasks for computation, communication, and storage, may be allocated to different units. Moreover, developments of hybrid quantum-classical algorithms are also critical for developing "killer applications" for near-term quantum computers. Proper integration of different quantum platforms and hybrid classical and quantum computing may provide significant advantages in real-world applications.

In the future, significant breakthroughs, such as fault-tolerant quantum operations, and quantum algorithms, will be achieved in medium-sized quantum systems. It is also desirable to achieve quantum advantage for applications in quantum chemistry, quantum machine learning, etc. Further developments could include specialized quantum machines, quantum clouds, and applications of quantum computing systems in quantum sensing and simulations. Finally, as quantum platforms develop toward fully scalable fault-tolerant quantum computing, we anticipate the emergence of broad real-world applications of quantum computing.

## ACKNOWLEDGMENTS

We thank Fei Yan for his contributions to the superconducting qubits section, and thank valuable discussions with Xiaodong He, we thank Chao-Yang Lu, Andrea Morello and Lieven M. K. Vandersypen for valuable comments, and we also thank Jiasheng Mai for figure polishing. This work was supported by the National Natural Science Foundation of China (Grants No. U1801661, 12174178, 11905098, 12204228, 12004165, 11875159, 12075110, 92065111, 12275117, 11905099, 11975117, 12004164, 62174076, 92165210, 11904157, 11661161018, 11927811, 12004371), National Key Research and Development Program of China (Grant No. 2019YFA0308100 and No. 2018YFA0306600), the Key-Area Research and Development Program of Guangdong Province (2018B030326001), the Guangdong Innovative and Entrepreneurial Research Team Program (2016ZT06D348, 2019ZT08C044), the Guangdong Provincial Key Laboratory (2019B121203002), the Guangdong Basic and Applied Basic Research Foundation (Grant No. 2021B1515020070 and No. 2022B1515020074), the Natural Science Foundation of Guangdong Province (2017B030308003), the Science, Technology and Innovation Commission of Shenzhen, Municipality (Grants No. KYTDPT20181011104202253, KQTD20210811090049034, K21547502, ZDSYS20190902092905285, KQTD20190929173815000, KQTD20200820113010023, JCYJ20200109140803865, JCYJ20170412152620376), Shenzhen Science and Technology Program (RCBS20200714114820298, RCYX20200714114522109), the Shenzhen-Hong Kong Cooperation Zone for Technology and Innovation (HZQB-KCZYB-2020050), the Anhui Initiative in Quantum Information Technologies (Grant No. AHY050000), the Innovation Program for Quantum Science and Technology (Grant No. 2021ZD0303205), Research Grants Council of Hong Kong (GRF No. 14308019), the Research Strategic Funding Scheme of The Chinese University of Hong Kong (No. 3133234). F.N. is supported in part by: Nippon Telegraph and Telephone Corporation (NTT) Research, the Japan Science and Technology Agency (JST) [via the Quantum Leap Flagship Program (Q-LEAP), and the Moonshot R&D Grant Number JPMJMS2061], the Japan Society for the Promotion of Science (JSPS) [via the Grants-in-Aid for Scientific Research (KAKENHI) Grant No. JP20H00134], the Asian Office of Aerospace Research and Development (AOARD) (via Grant No. FA2386-20-1-4069), and the Foundational Questions Institute Fund (FQXi) via Grant No. FQXi-IAF19-06.

**Author Contributions** M.-H.Y., Y.H., X.-H.D., J.L., D.L., B.-C.L., P.H., and Y.L. wrote the abstract and introduction. M.-H.Y. and B.C. wrote the quantum algorithms section. X.G., Y.Z., and F.N. wrote the superconducting qubits section. Y.L. wrote the trapped-ion qubits section. Y.H., P.H., and G.H. wrote the semiconductor spin qubits section. D.L. and C.Q. wrote the NV centers section. J.L., T.X., and X.-H.P. wrote the NMR system section. S.Y. wrote the neutral atom arrays section. H.W. wrote the photonic quantum computing section. X.-H.D., P.H., Y.H., and M.-H.Y. wrote the outlook and conclusion. The manuscript was revised by X.-H.D., P.H., J.Z., S.Z., F.N. and D.Y. with input from all other authors. D.Y. supervised the review project.